\documentclass[]{aa}
\UseRawInputEncoding 
\usepackage[utf8]{inputenc}
\usepackage{pdflscape}
\usepackage{xcolor}
\usepackage{lscape}
 \usepackage[hidelinks]{hyperref}
   \hypersetup{
      colorlinks = true,
      citecolor ={cyan}
   }

\usepackage{graphicx}

\newcommand{\msun}{$M_\odot$}
\newcommand{\kms}{km\,${\rm s}^{-1}$}

\newcommand{\HeII}{He\,{\sc ii}\,$\lambda 4686$}

\newcommand{\NVred}{N\,{\sc v}\,$\lambda 4945$}
\newcommand{\NVblue}{N\,{\sc v} $\lambda \lambda 4604, 4620$}

\newcommand{\NIV}{N\,{\sc iv}\,$\lambda 4058$}

\usepackage{txfonts}

\begin{document}

   \title{Constraints on the multiplicity of the most massive stars known: R136 a1, a2, a3, and c}

   \author{T.\ Shenar\inst{1, 2}
          \and H.\ Sana\inst{3} 
          \and  P.\ A.\ Crowther\inst{4} 
          \and K.\ A.\ Bostroem\inst{5}          
          \and L.\ Mahy\inst{6} 
          \and F.\ Najarro\inst{2} 
          \and L.\ Oskinova\inst{7} 
          \and A.\ A.\ C.\ Sander\inst{8} 
          }

   \institute{
   \inst{1}{Anton Pannekoek Institute for Astronomy, University of Amsterdam, Postbus 94249, 1090 GE Amsterdam, The Netherlands }; \email{T.Shenar@uva.nl} \\     
             \inst{2}{Departamento de Astrof\'isica, Centro de Astrobiolog\'ia (CSIC-INTA), Ctra.\ Torrej\'on a Ajalvir km 4, 28850 Torrej\'on de Ardoz, Spain}\\   
             \inst{3}{Institute of Astronomy, KU Leuven, Celestijnenlaan 200D, 3001 Leuven, Belgium}\\
             \inst{4}{Department of Physics \& Astronomy, Hounsfield Road, University of Sheffield, Sheffield, S3 7RH United Kingdom}\\
             \inst{5}{Steward Observatory, University of Arizona, 933 North Cherry Avenue, Tucson, AZ 85721-0065, USA}\\
             \inst{6}{Royal Observatory of Belgium, Avenue circulaire/Ringlaan 3, B-1180 Brussels, Belgium}\\
             \inst{7}{Institute for Physics and Astronomy, University Potsdam, D-14476 Potsdam, Germany}      
             \inst{8}{Zentrum f\"ur Astronomie der Universit\"at Heidelberg, Astronomisches Rechen-Institut, M\"onchhofstr. 12-14, 69120 Heidelberg}             
}

   \date{Received -; accepted -}


  \abstract
   {The upper stellar mass limit is a fundamental parameter for simulations of star formation, galactic chemical evolution, and stellar feedback. An empirical bound  on this parameter is therefore highly valuable. The most massive stars known to date are R\,136~a1, a2, a3, and c within the central cluster R\,136a of the Tarantula nebula in the Large Magellanic Cloud (LMC), with reported masses in excess of $150-200\,M_\odot$ and initial masses of up to $\approx 300\,M_\odot$. However, the mass estimation of these stars relies on the assumption that they are single.    }
   {Via multi-epoch spectroscopy, we provide for the first time constraints on the presence of close stellar companions to the most massive stars known for orbital periods of up to $\approx 10\,$yr.}
   {We collected three epochs of spectroscopy for R\,136~a1, a2, a3, and c with the Space Telescope Imaging Spectrograph (STIS) of the Hubble Space Telescope (HST)  in the years 2020-2021  to probe potential radial-velocity (RV) variations. We combine these epochs with an additional HST/STIS observation taken in 2012. For R\,136~c, we also use archival spectroscopy obtained with the Very Large Telescope (VLT). We use cross-correlation to quantify the RVs, and establish constraints on possible companions to these stars up to periods of $\approx 10\,$yr. Objects are classified as binaries when the peak-to-peak RV shifts exceed 50\,\kms, and when the RV shift is significant with respect to errors. }
   {R\,136~a1, a2, and a3 do not satisfy the binary criteria and are thus classified as putatively single, although formal peak-to-peak RV variability on the level 40\,\kms~is noted for a3. Only R\,136~c is classified as binary, in agreement with literature. We can generally rule out massive companions ($M_2 \gtrsim 50\,M_\odot$) to R\,136~a1, a2, and a3 out to orbital periods of $\lesssim 1\,$yr (separations $\lesssim 5\,$au) at 95\% confidence, or out to tens of years (separations $\lesssim 100$\,au) at 50\% confidence. Highly eccentric binaries ($e \gtrsim 0.9$) or twin companions with similar spectra could evade detection down to shorter periods ($\gtrsim 10\,$d), though their presence is not supported by the relative X-ray faintness of R\,136~a1, a2, and a3. We derive a preliminary orbital solution with a 17.2\,d period for the X-ray bright binary R\,136~c, though more data are needed to conclusively derive its orbit. }
   {Our study supports a lower bound of $150-200\,M_\odot$ on the upper-mass limit at LMC metallicity.}

   \titlerunning{Constraints on the multiplicity of the most massive stars known}
   \authorrunning{T. Shenar et al.}

   \maketitle
%
%
\section{Introduction}\label{sec:intro}

The upper mass limit of stars ($M_{\rm max}$) as a function of metallicity ($Z$) is one of the most fundamental parameters that dictate the properties of galaxies. This is because the ecology, energy budget, and integrated spectral appearance of galaxies are largely determined by the most massive stars they host \citep[$M\gtrsim 50\,M_\odot$;][]{Crowther2010, Doran2013, Ramachandran2019}. Moreover, the most massive stars are invoked in the context of a plethora of unique phenomena, from pair-instability supernovae and long-duration $\gamma$-ray bursts \citep{Fryer2001, Woosley2007, Langer2012, Smartt2009, Quimby2011, Fryer2001, Woosley2007} to the early chemical enrichment of globular clusters \citep{Gieles2018, Bastian2018}.  Establishing $M_{\rm max}$ from first principles or simulations of star formation is challenging due to a variety of uncertainties, and estimates vary from  $\approx 120$~\msun\ to a few $\approx 1000\,M_\odot$, depending on $Z$ and modelling assumptions \citep[e.g.][]{Larson1971, Oey2005, Figer2005}. It is therefore essential to identify and weigh the most massive stars in our Galaxy and nearby lower-metallicity galaxies such as the Small and Large Magellanic Clouds (SMC, LMC).

Stars initially more massive than ${\approx} 100\,M_\odot$, dubbed very massive stars (VMS), tend to have emission-line dominated spectra stemming from their powerful stellar winds already on the main sequence.  Such stars spectroscopically appear as Wolf-Rayet (WR) stars \citep{deKoter1997}. Being N-rich owing to the CNO burning cycle, they belong to the nitrogen WR sequence (WN). Unlike classical WR stars, which are evolved and H-depleted massive stars, VMSs typically show substantial surface hydrogen mass fractions ($X_{\rm H} \gtrsim 40\%$), and are usually classified as WNh to indicate a H-rich atmosphere.

In the case of  double-lined spectroscopic binaries (SB2),  the mass ratio and minimum masses of both components  ($M_{1, 2} \sin^3 i$, where $i$ is the orbital inclination)  can be established via Newtonian mechanics. If the inclination is known additionally, then the true masses can be derived. Best constraints are obtained for eclipsing binaries, such as the massive Galactic binaries \object{WR 43a} \citep[][$M_1 = 116\pm31\,M_\odot$, $M_2 = 89\pm16\,M_\odot$]{Schnurr2008}, \object{WR~21a} \citep[][$M_1 = 93\pm2\,M_\odot$, $M_2 = 53\pm1\,M_\odot$]{Tramper2016, Barba2022}, \object{WR~20a} \citep[][$M_1 = 83\pm5\,M_\odot$, $M_2 = 82\pm5\,M_\odot$]{Rauw2004, Bonanos2004}, and the SMC binary \object{HD\,5980}, which hosts a luminous blue variable (LBV) and a WR star of masses $60-70\,M_\odot$ \citep{Koenigsberger2014}.  The inclination can also be constrained from interferometry \citep[e.g.,][]{Richardson2016, Thomas2021} or from spatially resolved structures, such as the Homunculus nebula of of the Galactic binary \object{$\eta$\,Car}, a luminous LBV+WR system with masses $\approx 100 + 60\,M_\odot$ \citep{Madura2012, Strawn2023}.  Alternatively, the inclination can be constrained via polarimetry \citep{Brown1978, Robert1992} or wind eclipses \citep{Lamontagne1996}, as was the case for the LMC binaries  \object{R\,144} (alias \object{BAT99 118}) and R\,145 (alias \object{BAT99 119}), which host similar-mass components with current masses of $\approx 70-80\,M_\odot$ and initial masses of $\approx 150\,M_\odot$ \citep{Shenar2017b, Shenar2021}.

When the inclination cannot be measured, the mass ratio and minimum masses provide nevertheless important parameters, which, in conjunction with other methods, constrain the true masses of the components.  Examples include the the LMC colliding-wind binary \object{Melnick~33Na} \citep[][$M_1 =83\pm19\,M_\odot$, $M_2 =48\pm11\,M_\odot$]{Bestenlehner2022}, and the most massive binary known to date, \object{Melnick~34} (alias \object{\mbox{BAT99}~116}), with derived component masses of $M_1 = 139\pm20\,M_\odot$ and $M_2 = 127\pm17\,M_\odot$~\citep{Tehrani2019}.

In the absence of a companion, the mass of a star is estimated by matching the derived stellar properties (mainly the luminosity $L$, effective temperature $T_{\rm eff}$, and $X_{\rm H}$) with structure or evolution models, yielding the evolutionary current and initial masses. It is important to note that such mass estimates assume an internal structure for the star (e.g., H-burning, He-burning), which is not always trivial  for WNh stars \citep[e.g., the case of R\,144,][]{Shenar2021}. Prominent examples for putatively single VMS and WNh stars  include the LMC WN5h star \object{VFTS 682} \citep{Bestenlehner2014} and several very massive WNh and Of stars in the Galactic clusters Arches \citep[e.g.][]{Figer2002, Najarro2004, Martins2008, Lohr2018} and \object{NGC~3603} \citep{Crowther2010}.  The latter method was also used to establish the masses of the most massive stars known, which are the subject of this study. These stars, which are classified WN5h and which reside in the dense central cluster R\,136 of the Tarantula nebula in the LMC, include
\object{R\,136~a1}\footnote{\object{RMC136 a1} on SIMBAD} (alias \object{\mbox{BAT99}~108}, $M = 200-300\,M_\odot$, a1 thereafter), \object{R\,136~a2} (alias \object{BAT99 109}, $M = 150-250\,M_\odot$, a2 thereafter), \object{R\,136~a3} (alias \object{BAT99 106}, $M = 150-200\,M_\odot$, a3 thereafter), and \object{R136~c} (alias \object{\mbox{BAT99}~112}, \object{VFTS~1025}, $M = 150-200\,M_\odot$, c thereafter). 

Earlier studies in the 1980s identified the central region of R\,136 as a single star with a mass $\gtrsim 1000\,M_\odot$ \citep{Cassinelli1981, Savage1983}, but later investigations with speckle interferometry  and the Hubble Space Telescope (HST) showed that the central region comprised distinct stellar sources, including a1, a2, and a3 \citep{Weigelt1985, Lattanzi1994, Hunter1995}. 
First mass measurements of these objects yielded masses of the order of $100\,M_\odot$  \citep[e.g.,][]{Heap1994, deKoter1997, Massey1998, Crowther1998}. However, \citet{Crowther2010} reported masses in excess of $200\,M_\odot$  using modern model atmospheres which account for iron line blanketing. Since then, the masses of a1, a2, a3, and c have been subject to  several revisions \citep{Hainich2014, Crowther2016,  RubioDiez2017, Bestenlehner2020, Brands2022}, but remain record-breaking in terms of current and initial masses.  A visual companion to a1 was identified by \citet{Lattanzi1994} with the HST's Fine Guidance Sensor (FGS),  and later with HST imaging by \citet{Hunter1995}. Recently, \citet{Khorrami2017} and  \citet{Kalari2022} confirmed the presence of this companion, and  detected another faint companion to a3 via speckle imaging. Accounting for these companions potentially lowers the mass estimates of a1 and a3 by $10-20\%$.

A drawback for mass estimates of putatively single stars is the assumption that they are single.  The presence of a contaminating companion could substantially alter the derived stellar parameters (especially $L$), and, in turn, the stellar masses.  The realisation that the majority of massive stars reside in binary systems \citep{Sana2012, Sana2013b} forces us to consider that R\,136~a1, a2, a3, and c may be members of such binaries.
In fact, relying on K-band spectroscopy acquired  over 22\,d, \citet{Schnurr2009} already identified R\,136~c as a potential binary with a 8.2\,d period. The relatively high X-ray luminosity ($\approx 10^{35}\,{\rm erg}\,{\rm s}^{-1}$) of R\,136~c  \citep{PZ2002, Townsley2006, Guerrero2008, Crowther2022} suggests that it is a colliding-wind binary, where both companions possess a fast wind, although a compact object companion is also a viable option.  

Thus far, a1, a2, and a3 were not probed for multiplicity for periods longer than a few weeks. 
In this study, we present results from a 1.5\,yr spectroscopic monitoring of R\,136~a1, a2, a3, and c obtained with the Hubble Space Telescope (HST), combined with a previous epoch obtained in 2012 and archival data for R\,136~c. By deriving the radial velocities (RVs) of the stars, we place constraints on potential companions.  We additionally derive a new orbital solution for R\,136~c. The reduction of the data is described in Sect.\,\ref{sec:data}, and their analysis is described in Sect.\,\ref{sec:analysis}. We discuss our results in Sect.\,\ref{sec:discussion} and provide a brief summary in Sect.\,\ref{sec:summary}.

\section{Data and reduction}\label{sec:data}

\begin{figure}
\centering
\includegraphics[width=.5\textwidth]{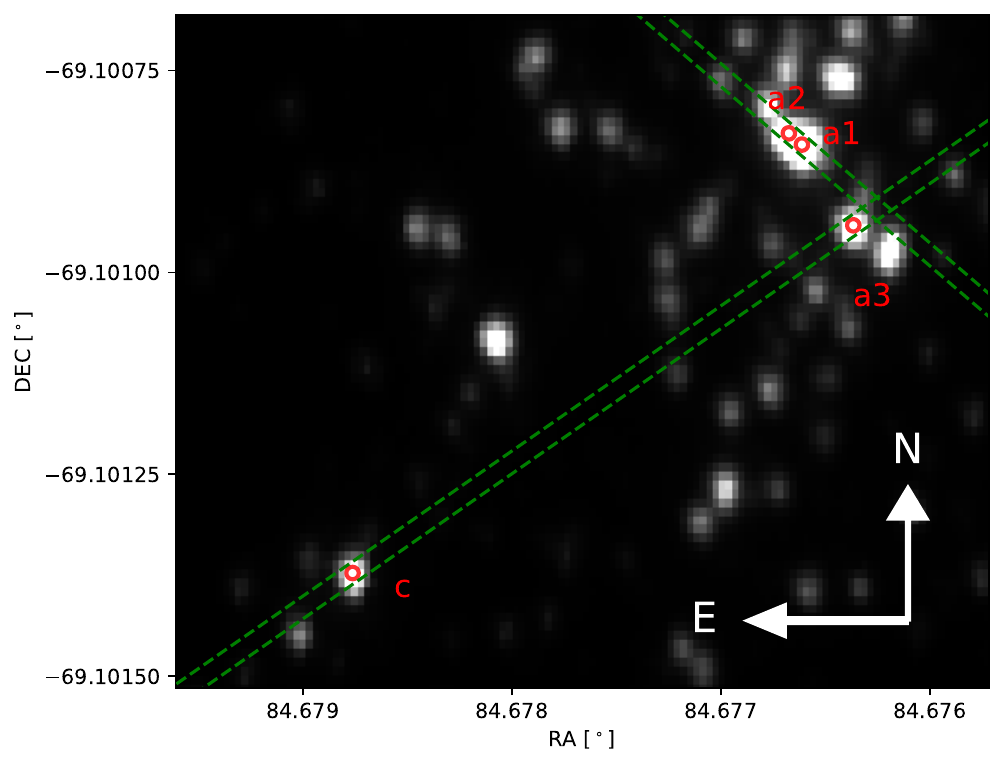}
\caption{Image of the central 3'' $\times$ 5'' ($\approx 0.7 \times 1.2$ parsec) core of R\,136 taken in 2009 with the HST's Wide Field Camera 3 (WFC3) in the ultraviolet and visible light (UVIS) channel with the F555W filter ($\lambda_0 \approx 5500\,$\AA, proposal ID: 11360).  Marked are the positions of our four targets and the two slit positions used to acquire the data. } 
\label{fig:SlitPos}
\end{figure}

Our investigation relies primarily on three epochs of spectroscopy obtained with the Space Telescope Imaging Spectrograph (STIS) mounted on the HST (PI: Shenar, proposal ID: 15942). We used a slit aperture of  $52''\times0.1''$ with the G430M filter at the central wavelengths 3936 (3770-4101\AA), 4706 (4540 - 4872\AA), and 4961 (4795 - 5127\AA). Each pair of stars (a1, a2), (a3, c) defined a STIS slit positioning to enable the acquisition of the spectra of two stars during a single pointing (Fig.\,\ref{fig:SlitPos}). The resulting position angles (PA) are 106$^\circ$ (or 286$^\circ$) and 162.2$^\circ$ (or 342.2$^\circ$) for the pairs (a1, a2) and (a3, c), respectively. The PA of the pair (a1, a2) is similar to the PA used by \citet{Crowther2016} in their scanning of the R\,136 cluster with STIS (where the PA was 109$^\circ$ or 289$^\circ$).
Overall, three epochs of observations were acquired on 28 March 2020 (MJD 58936.20), 28 September 2020 (MJD 59120.07), and 14 September 2021 (MJD 59471.74) for the pair (a1, a2), and on  25 May 2020 (MJD 56023.42), 26 November 2020 (MJD 58994.06), and 14 May 2021 (MJD 59179.31) for the pair (a3, c)  for each of the three spectral bands. Each exposure was divided into two dithered subexposures for removal of cosmics. The G430M filter has a spatial dispersion of 0.05''/pixel The signal-to-noise ratio (S/N) of the data is $\approx 20-80$ per pixel, depending on the star and the spectral domain. The spectral resolving power is $R = \lambda / \Delta \lambda \approx 6\,000$  with a dispersion of $\Delta \lambda = 0.28\AA$. 

\begin{figure}
\centering
\includegraphics[width=.5\textwidth]{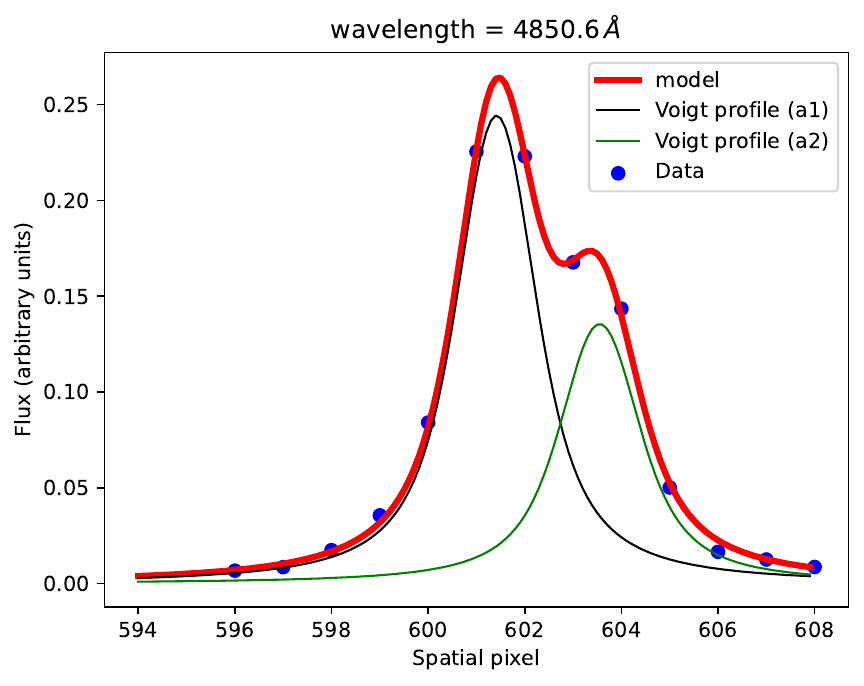}
\caption{Fit of two Voigt profiles representing the PSFs of a1 and a2 to the  flux across the cross-dispersion axis  for the  28 March 2022 epoch at 4850.6$\AA$. The Voigt profiles are used to compute the relative weight of each datapoint to the flux of each star at each given wavelength.}
\label{fig:Voigt}
\end{figure}

\begin{figure}
\centering
\includegraphics[width=.5\textwidth]{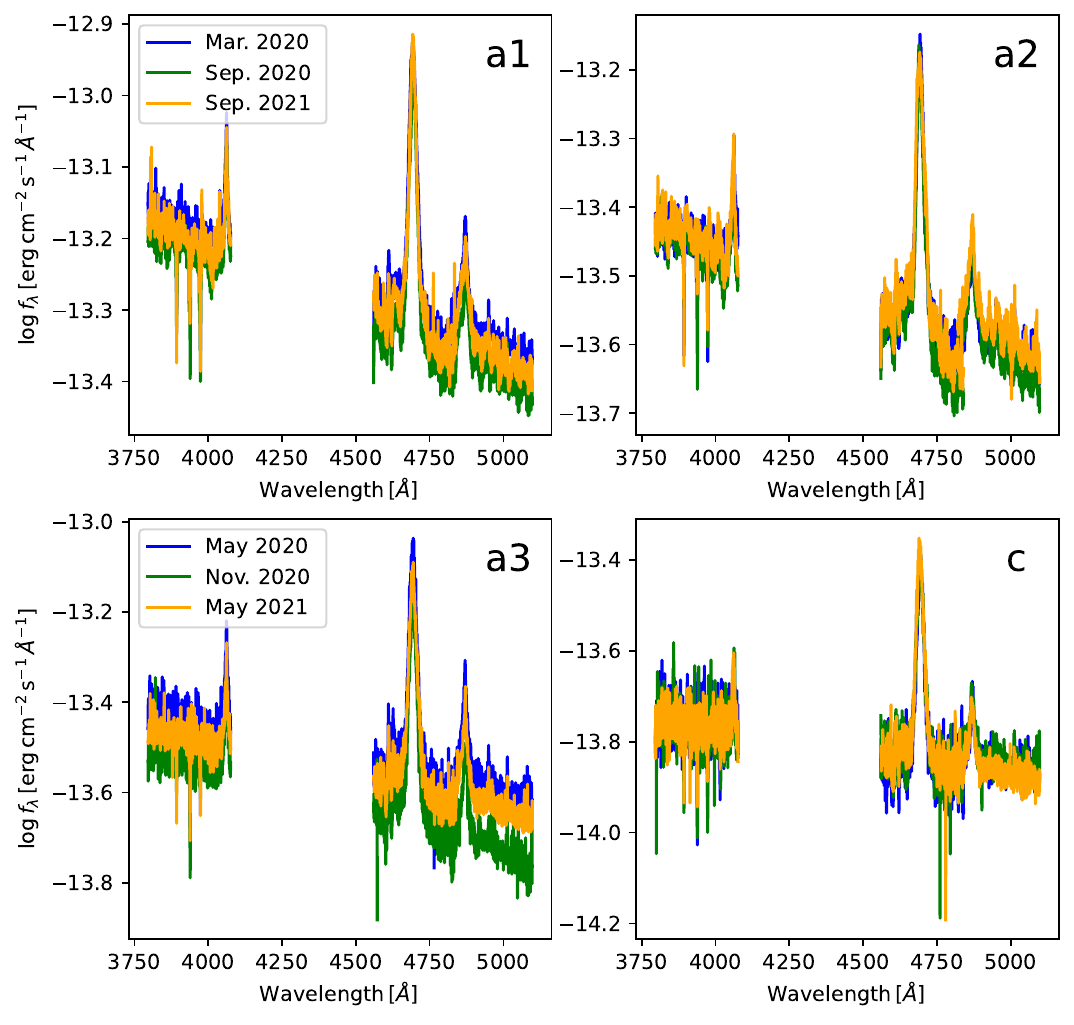}
\caption{Extracted calibrated fluxes for a1, a2, a3, and c, for the new epochs presented here (see labels and legends).   } 
\label{fig:FluxPlot_lambda}
\end{figure}

The extraction of the spectra of the stars R\,136~a3 and c across the slit is straight forward, since they are well separated spatially. The extraction of the pair (a1, a2), however, is less trivial, since the point spread functions (PSFs) of the two sources overlap (see Fig.\,\ref{fig:SlitPos}). To extract the spectra, 
we fit Voigt profiles with identical width parameters (to mimic the PSF) to the flux across the cross-dispersion direction, as shown in Fig.\,\ref{fig:Voigt}.  The fitting of the Voigt profile is performed in a wavelength-dependent fashion, such that the flux across the spatial direction was fit for each wavelength bin. We fix the separation between the Voigt profiles to 0.113'' (or 2.25 HST pixels), as found by \citet{Kalari2022}, and fix the amplitude ratios of the Voigt profiles to the magnitude ratio derived by \citet{Kalari2022}. Avoiding the latter resulted in an instrumental wavy pattern that compromised the RV measurements. We fit for the Voigt broadening parameters $\sigma, \gamma$ as a function of wavelength, but enforce both Voigt profiles of a1 and a2 to share the same parameters.  The resulting spectral energy distributions are shown in Fig.\,\ref{fig:FluxPlot_lambda}. 

The flux levels of the four stars are relatively consistent in the three available epochs, though $\approx$10\% variations are seen in a1 and a3. Such discrepancies are typical for the narrow-slit mode of STIS, which does not fully account for slit losses \citep[e.g.][]{Lennon2021}. The overall flux level is consistent between the different epochs and is in agreement with the flux level presented by \citet{Crowther2010}. However, we cannot rule out some contamination between a1 and a2 for strong lines such as \HeII~(see below), since in this case the PSFs are not well resolved. Results obtained for the \HeII~line for these components should be therefore taken with caution.

A similar technique was used by \citet{Crowther2016} for their analysis of STIS spectroscopy of the R\,136 cluster. The flux-calibrated spectra show a general agreement with those presented by \citet{Crowther2016}, although the underlying spectral energy distribution for a1 and a2 depends on the PA, suggesting that the flux variability observed between the epochs is linked to the observational setup rather than intrinsic. For our study, only normalised spectra were used.  The extracted spectra were rectified using a homogeneous set of pre-selected continuum points.

To verify our extraction methodology, we newly extracted the spectra of a1, a2, and a3 from observations acquired in 2012 with STIS which were extracted and analysed by \citet{Crowther2016} using the {\sc multispec} package \citep{Maiz-Apellaniz2005, Knigge2008}. The extractions match well with each other. The 2012 spectra for a1, a2, and c are combined with the newly acquired 2020-2021 spectra in our investigation to boost the binary detection probability.
We also inspected the interstellar Ca\,{\sc ii} K and H lines at 3934.77$\AA$ and 3969.59$\AA$ (wavelengths in vacuum) as a check on the absolute wavelength calibration of the spectra.

\begin{figure*}
\centering
\includegraphics[width=\textwidth]{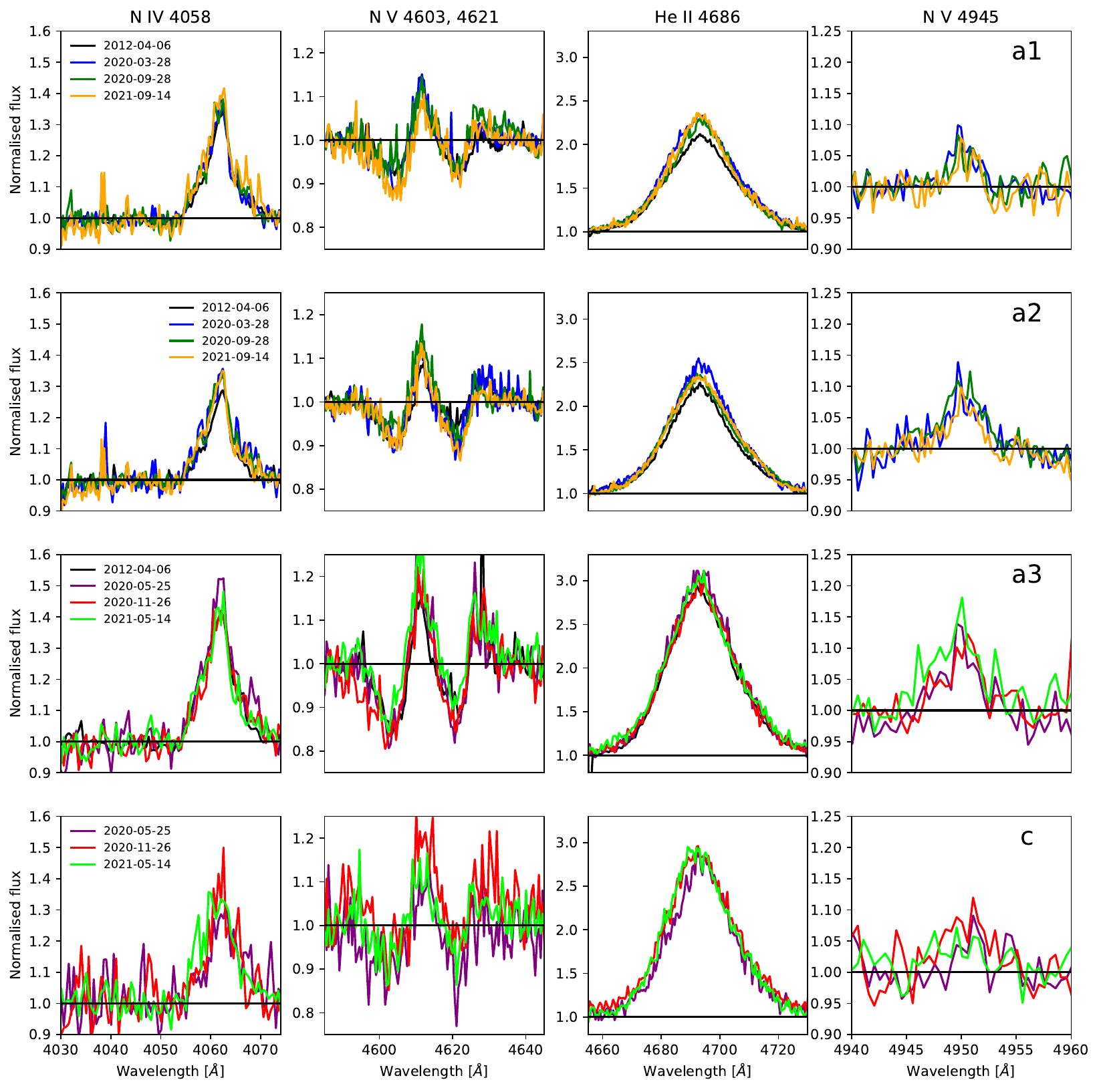}
\caption{Extracted STIS spectra for a1, a2, a3, and c (from top to bottom), focusing on the diagnostic lines \NIV, \NVblue, \HeII, and \NVred~(from left to right). Epochs are listed in the legends. The spectra of a3 and c are binned at $\Delta \lambda = 0.5\,\AA$ for clarity. } 
\label{fig:specs}
\end{figure*}

The extracted spectra (both the new data as well as the original 2012 epochs by \citealt{Crowther2016}) are shown in Fig.\,\ref{fig:specs}. The spectra cover only few features that could correspond to cooler companions (e.g., He\,{\sc i}\,$\lambda 4713$), but these features appear flat for R\,136~a1, a2, and a3 (for c, see Sect.\,\ref{subsec:ARGUS}), in agreement with previous studies of these objects.  We therefore only show the main diagnostic lines: \NIV, \NVblue, \HeII, and \NVred. 

Before advancing to the analysis, it is interesting to already note that no clear indications for RV variability are seen from an inspection of Fig.\,\ref{fig:specs}, with the exception of star c. 
Identifying companions on the basis of spectral appearance (as opposed to RV variability) is not viable here. Companions of interest in this study would have masses $\gtrsim 50\,M_\odot$, and such companions show primarily H and He\,{\sc ii} lines, which overlap with those of the WR primaries. Searching for RV variability is hence the method of choice with the available data.

For R\,136~c, we use  34 archival spectra in addition to the HST data.  These data cover five observing epochs (PI: Evans, ID: 182.D-0222) and were acquired in 2008-2010 with the Fibre Large Array Multi Element Spectrograph (FLAMES) ARGUS integral field unit (IFU) mounted on UT2 of the Very Large Telescope (VLT). Each spaxel of ARGUS spatially covers 0.52''.  The spectra cover the range $3960-4570\,\AA$ with a resolving power of $R = 10\,500$, a dispersion of $\Delta \lambda = 0.2\,\AA$, and a typical S/N of 50-100 per pixel. The retrieval and reduction of the data are described in \citet{Evans2011}. In addition, we retrieved a single spectroscopic observation acquired in 2001 with the Ultraviolet and Visual Echelle Spectrograph (UVES) mounted on UT2 of the VLT. We only use the spectrum covering the range 3700-5000\,\AA, which includes the \NIV~line. The spectrum has a resolving power of $R=40\,000$ and a S/N of $\approx 20$ per pixel, with a dispersion of $\Delta \lambda = 0.015\,\AA$.
The data are described in \citet{Cox2005}, and are retrieved in reduced form from the European Southern Observatory's (ESO) archive. We ensure wavelength calibration to within a few \kms~using the Ca\,{\sc ii} H line at 3969.59$\AA$, which is present in the HST, ARGUS, and UVES datasets.


\section{Analysis}\label{sec:analysis}

\subsection{Cross correlation}
\label{subsec:HSTRVs}

\begin{figure}
\centering
\includegraphics[width=.5\textwidth]{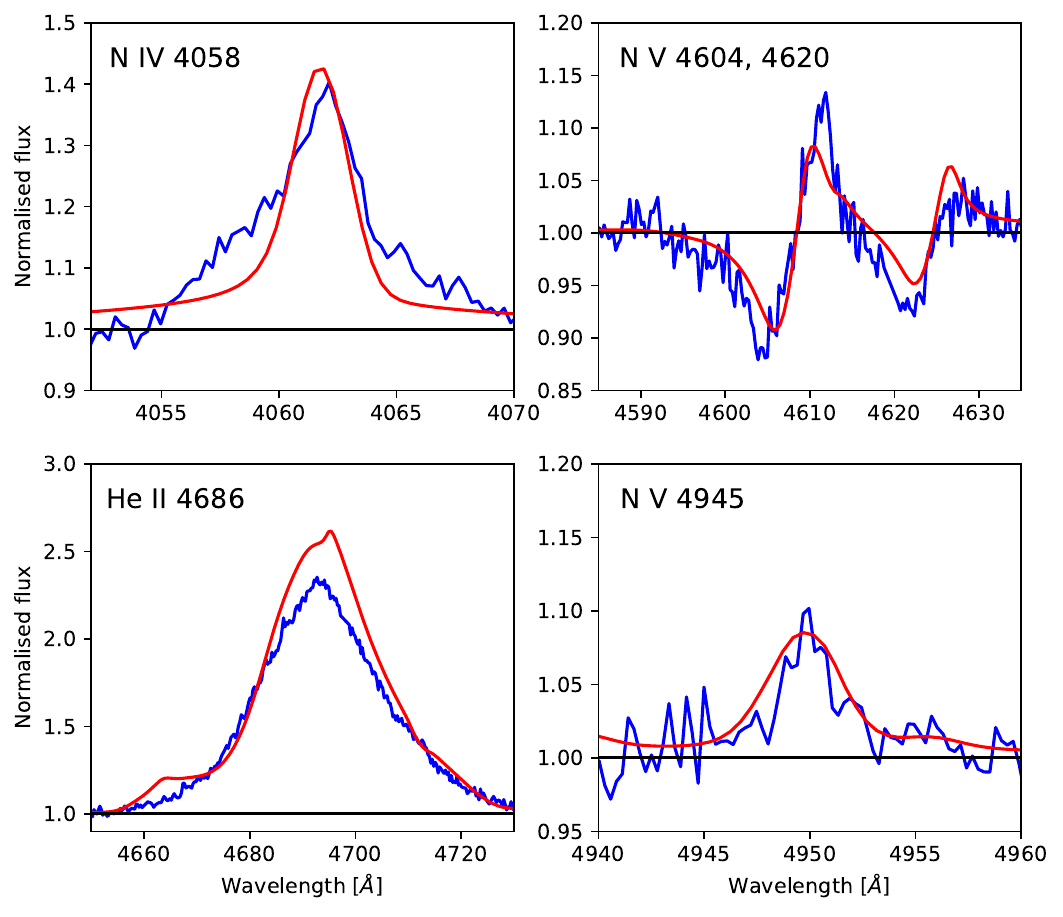}
\caption{Comparison between the synthetic PoWR model (solid red line) used for cross-correlation and the normalised spectrum of a1 taken in March 2020 (noisy blue line). The line profiles of the PoWR model were shifted by 330\,\kms~and the equivalent widths of the \NIV~and \NVred~profiles were scaled to the observations for the sake of plotting (the equivalent width has no impact on the cross-correlation algorithm). The spectra of a2, a3, and c are comparable, and so is the match between the model and the data.  } 
\label{fig:PoWRComp}
\end{figure}

The main tool with which our targets are probed for multiplicity is the measurement of RVs via maximisation of cross-correlation functions (CCF). The technique is described by \citet{Zucker1994}, and is frequently implemented for WR binaries \citep[e.g.,][]{Shenar2017b, Shenar2019, Shenar2021, Dsilva2020, Dsilva2022, Dsilva2023}. Briefly, the CCF is computed in a particular spectral range (or multiple ranges) as a function of Doppler shift using a pre-specified template that should represent the star. While for WR stars the template is usually produced by co-adding the individual observations, this does not yield satisfactory results in our case due to the limited number of observations and the modest S/N. Moreover, we refrain from cross-correlating multiple lines simultaneously, since lines in WR spectra are formed in different radial layers and therefore often produce systematic shifts with respect to one another \citep[e.g.][]{Shenar2017}. Instead, we use a synthetic spectrum computed with the Potsdam Wolf-Rayet (PoWR) model atmosphere code \citep{Hamann2003, Sander2015} tailored for the analysis of these objects by \citet{Hainich2014}. While this yields absolute RVs, an absolute RV calibration for WR stars is highly model dependent since the line centroids sensitively depend on the atmosphere parameters. However, this does not impact our study, since the detection of binaries relies on relative RVs. A comparison between the PoWR model used here and the spectrum of a1 is shown in Fig.\,\ref{fig:PoWRComp}.

The results from the CCF analysis for \NIV, \NVblue, \HeII, and \NVred~are shown in Fig.\,\ref{fig:DeltaRV}. Tables\,\ref{tab:RVs} and \ref{tab:EWs} in Appendix\,\ref{appendix:log}  provide a compilation of these measurements and  the measured EWs of these lines. Upper bounds on the statistical errors on the EWs are computed as in \citet{Chalabaev1983}.  Significant EW variability is noted between the 2012 epoch and the other epochs in the \HeII~line belonging to a1. While such variability is not untypical for WR stars \citep[e.g.,][]{Moffat1992, Lepine1999}, this result could also be spurious given the difficulty in the more challenging extraction of the \HeII~line (see Sect.\,\ref{sec:data}.

While the nitrogen lines are typically considered as the best RV probes of WN stars as they form relatively close to the stellar surface, the modest S/N becomes a limiting factor. In this context, the \HeII~line  offers an important high S/N RV probe, but its interpretation should be treated with caution for a1 and a2 given possible cross-contamination in this line. The fact that the RVs of a1 and a2 from the \NIV, \NVblue, and \NVred~lines are consistent between the epochs, but the those of the \HeII~line are strongly variable, suggests that this RV variability is not genuine.

\begin{figure}
\centering
\includegraphics[width=.5\textwidth]{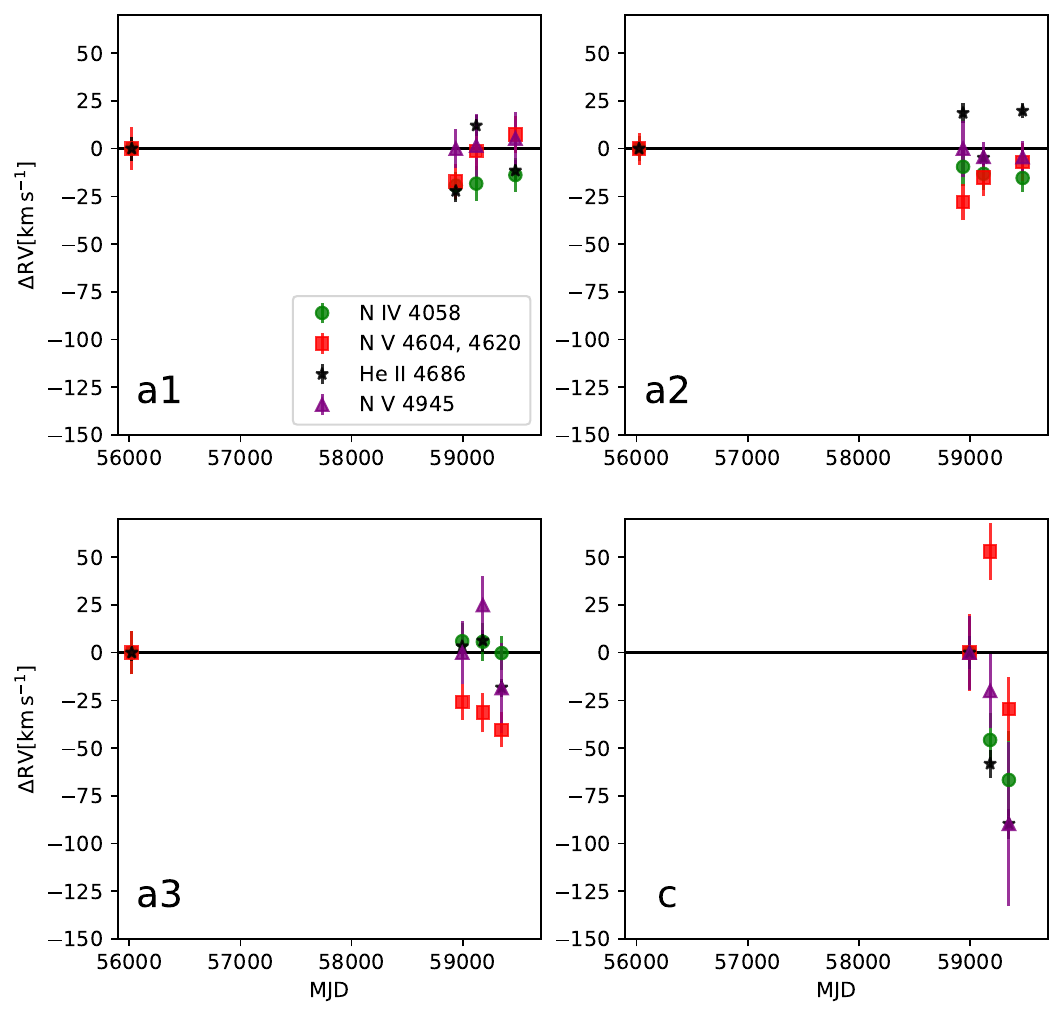}
\caption{Relative RVs measured for the four diagnostic spectral features in the spectra (when available; see legend). The first RV is always calibrated to zero. Only component c is classified as binary by fulfilling Eqs.\,(\ref{eq:p2p}) and  (\ref{eq:Dpp}).} 
\label{fig:DeltaRV}
\end{figure}

In principle, to classify RV variables into binaries vs.\ putatively single, one commonly adopts a significance criteria on the peak-to-peak RV variability (see, e.g., \citealt{Sana2013b}). Due to the intrinsic variability of WR stars, it is often not trivial to find a single criterion. A significance criterion on the peak-to-peak (pp) RV shift which is commonly adopted is

\begin{equation}
   \Delta  {\rm RV}_{\rm pp, \sigma} =   \underset{i \neq j}{\rm max} \left\{ \frac{ \left| {\rm RV}_i - {\rm RV}_j \right| }{\sqrt{\sigma_i^2 + \sigma_j^2} } \right\}  > 4,
\label{eq:p2p}
\end{equation}
where $\sigma$ is the corresponding error on the RV measurement. The threshold 4 is considered conservative, resulting in a false-positive probability of roughly 0.1\% \citep{Sana2013b}. However,  because of the intrinsic variability of WR stars, $\sigma$ may be underestimated,  and this criterion alone can lose validity. Hence, in addition, we also invoke a threshold criterion on the peak-to-peak RV difference, $\Delta {\rm RV}_{\rm pp} ={\rm max} \{{\rm RV}_i - {\rm RV}_j\}$. \citet{Dsilva2023} conducted an RV monitoring survey of 11 late-type WN stars of spectral classes similar to those of a1, a2, a3, and c, and found the threshold $\Delta {\rm RV}_{\rm pp} > 50\,$\kms~clearly separated spectroscopic binaries from potentially single stars, and that intrinsic variability can lead to apparent RV variations of up to $50\,$\kms, depending on the wind and stellar properties. Our second criterion for a binary classification is thus:

\begin{equation}
    \Delta  {\rm RV}_{\rm pp} = \underset{i \neq j}{\rm max} \left| {\rm RV}_i - {\rm RV}_j  \right| > 50\,{\rm km}\,{\rm s}^{-1}
\label{eq:Dpp}
\end{equation}

While the choice of this threshold impacts our classification to binary or single, the bias discussion provided in Sect.\,\ref{sec:discussion} addresses this issue.  Table\,\ref{tab:BinStat} summarises whether or not Eqs.\,(\ref{eq:p2p}) and (\ref{eq:Dpp}) are fulfilled for each of the spectral diagnostics. When both conditions are satisfied, we flag the star as binary. Evidently, the only star that satisfies both conditions is R\,136~c (with the \HeII~line, and marginally with the nitrogen lines). In contrast, a1, a2, and a3 do not satisfy both conditions for any of the lines, and are hence flagged as putative single. The fact that R\,136~c is classified as binary is consistent with the findings of \citet{Schnurr2009}, who derived a tentative 8.2\,d orbital period for this object, and its high X-ray luminosity, suggestive of colliding winds or a compact object in the binary \citep{PZ2002, Crowther2022}.

\begin{table} 
\centering
\setlength{\tabcolsep}{3.2pt}
\label{tab:BinStat}
\small
\centering 
\caption{Binary status of R\,136~a1, a2, a3, and c}
\begin{tabular}{llcccc} \hline \hline
Object & Condition & N\,{\sc iv}\,$4058$ & N\,{\sc v}\,$ 4603, 4621$  & He\,{\sc ii}\,$ 4686$ & N\,{\sc v}\,$ 4945$ \\
\hline
a1 & Eq.\,(\ref{eq:p2p}) & no & no & yes $^{\rm (a)}$ & no \\ 
a1 & Eq.\,(\ref{eq:Dpp}) & no & no & no & no \\ 
a1 & binary? & no & no & no & no \\ 
\hline
a2 & Eq.\,(\ref{eq:p2p}) & no & no & yes $^{\rm (a)}$ & no \\ 
a2 & Eq.\,(\ref{eq:Dpp})  & no & no & no & no \\ 
a2 & binary? & no & no & no & no \\ 
\hline
a3 & Eq.\,(\ref{eq:p2p}) & no & no & no & no \\ 
a3 & Eq.\,(\ref{eq:Dpp}) & no & no & no & no \\ 
a3 & binary?  & no & no & no & no \\ 
\hline
c & Eq.\,(\ref{eq:p2p})  & no  & no  & yes       & no \\ 
c &  Eq.\,(\ref{eq:Dpp}) & yes & yes & yes       & yes \\ 
c & binary?              & no  & no  & {\bf yes} & no \\ 
\hline
\end{tabular}
\tablefoot{$^{\rm (a)}$ This result should be taken with caution, since contamination between a1 and a2 in the strong \HeII~line is possible due to  blending of the PSFs.} 
\end{table}

Before advancing to the interpretation of these results,  one may wonder whether a 1D CCF method is valid if these objects are double-lined spectroscopic binaries (SB2). Given the spectral appearance of our targets, the only companions which could be relevant in terms of contributing sufficient flux to bias the results are O-type stars or WR stars.  O-type dwarfs typically have absorption-line dominated spectra with weak to non-existing features belonging to N\,{\sc iv} or N\,{\sc v}, and much weaker features in the \HeII~line compared to a WR star \citep{Walborn1990, Walborn2002}. A contamination with an O dwarf therefore poses no danger to our RV measurement methodology. 

\begin{figure}
\centering
\includegraphics[width=.5\textwidth]{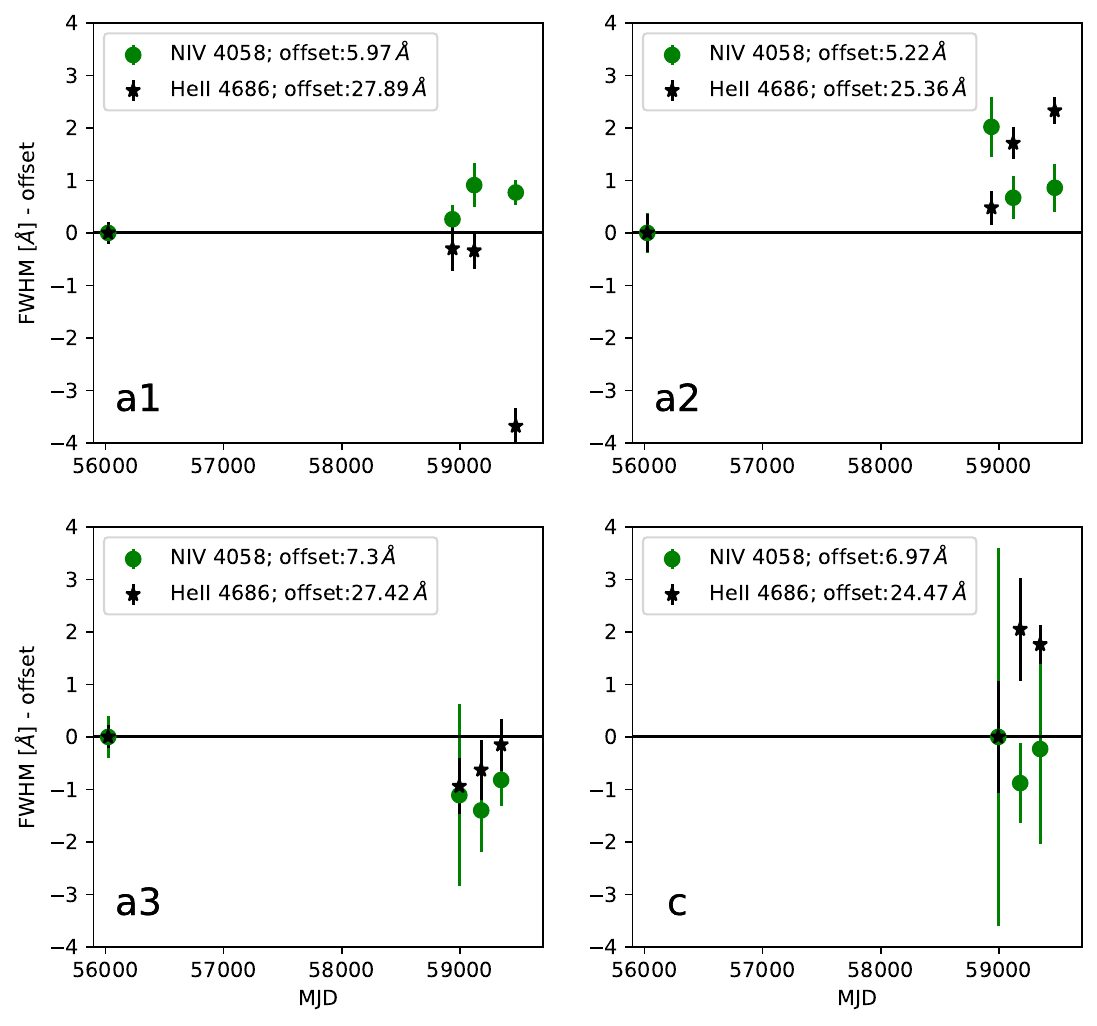}
\caption{FWHMs of the \NIV~and \HeII~lines for components a1, a2, a3, and c.  The values are offset relative to the FWHMs measured in the first available epoch for each star. The offset values are provided in the legend. Only the FWHMs of \HeII~for component c are significantly variable (on a 4$\sigma$ level). } 
\label{fig:FWHM}
\end{figure}

However, a contamination with an Of star, a transition O/WR star \citep{Crowther2011}, or a WN star could impact our interpretation. For example, consider the case of two WN stars with similar \NIV~or \HeII~line profiles. Typical peak-to-peak RV amplitudes may fall below the full-width half-maximum (FWHM) of these lines, implying that the line profiles would remain blended and show a marginal or even vanishing RV shift \citep[e.g.,][]{Sana2011}.  The same argument holds for P-Cygni lines (such as \NVblue), although the effect is more difficult to quantify. Thus, instead of an RV variation, one would observe a periodic change in the FWHM of the line. Excess emission stemming from wind-wind collisions (WWC) may also be added to this line, further changing its profile \citep{Luehrs1997}. For this reason, we also measured the FWHMs of the \NIV~and \HeII~lines for the a1, a2, a3, and c (Fig.\,\ref{fig:FWHM}). To obtain the FWHMs and their respective errors, we fit Gaussian profiles to the \NIV~and \HeII~lines and generate 1000 spectra with the same underlying Gaussian profile and S/N of the original data. The FWHM is then taken as the average of the FWHMs of these 1000 simulated spectra, and the error is their standard deviation. The results are shown in Fig.\,\ref{fig:FWHM}, and are also provided in Table\,\ref{tab:FWHM}. Neither of the stars exhibits strong variability, with only the \HeII~line of R\,136~c being significantly variable (on a 4$\sigma$ level). The interpretation of these results will be conducted in Sect.\,\ref{sec:discussion}.

\subsection{Orbital analysis of R\,136~c}
\label{subsec:ARGUS}

We follow a similar analysis methodology using the 34 calibrated FLAMES spectra and single UVES spectrum available for R\,136~c, which probe six distinct observational epochs in addition to the three HST epochs. The only robust RV probe in the available spectral range is the \NIV~line, for which the same PoWR template is used as in Sect.\,\ref{subsec:HSTRVs}.  The full list of RVs is available in Tables\,\ref{tab:RVs} and \ref{tab:RVsC}. The amplitude of the RV variability is in apparent agreement with the HST data.

\begin{figure}
\centering
\includegraphics[width=.5\textwidth]{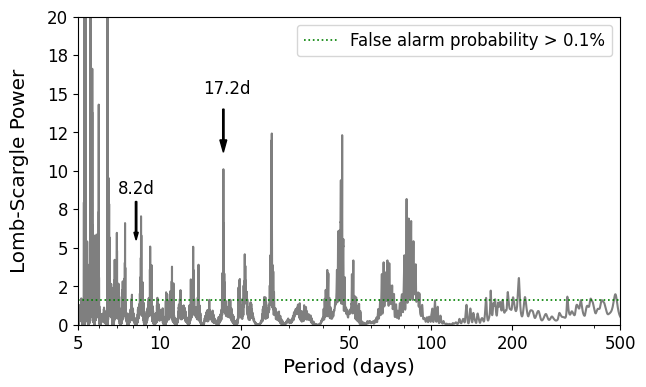}
\caption{Lomb-Scargle periodogram of the full RV set derived from the HST + FLAMES + UVES data for R\,136~c. Our favoured period of 17.2\,d is marked, along with the previously published 8.2\,d period, which is not supported by this study.  } 
\label{fig:LSPeriod}
\end{figure}

Figure\,\ref{fig:LSPeriod} shows a Lomb-Scargle periodogram derived for the full set of RVs (HST + FLAMES + UVES). Evidently, multiple peaks are present in the periodogram, with the most prominent peaks at $P = 5.277, 5.345, 5.563, 17.20,  25.94, 47.15$, and $83.06\,$d. For this set of periods, we use Python's lmfit minimisation package \footnote{https://lmfit.github.io/lmfit-py} with the differential evolution method to constrain the time of periastron $T_0$, systemic velocity $V_0$, RV semi amplitude $K_1$, argument of periastron $\omega$, and eccentricity $e$ via 

\begin{equation}
    {\rm RV}(\nu) = V_0 + K_1\,\left( \cos \nu + e\,\cos \omega \right).
\label{eq:Kepler}
\end{equation}

\begin{figure}
\centering
\includegraphics[width=.5\textwidth]{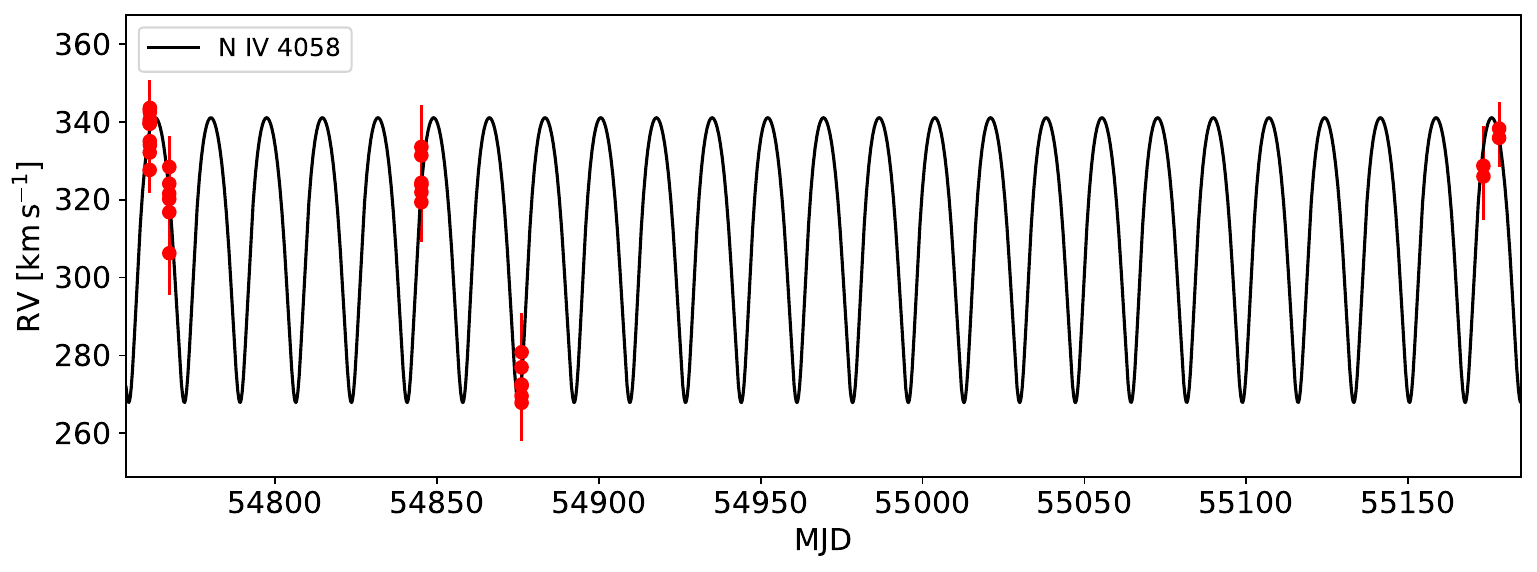}
\caption{The RVs measured for the FLAMES/ARGUS data for R\,136~c as a function of MJD, compared to the best-fitting RV curve corresponding to $P=17.2051\,$d.} 
\label{fig:OrbitTime}
\end{figure}

\begin{figure}
\centering
\includegraphics[width=.5\textwidth]{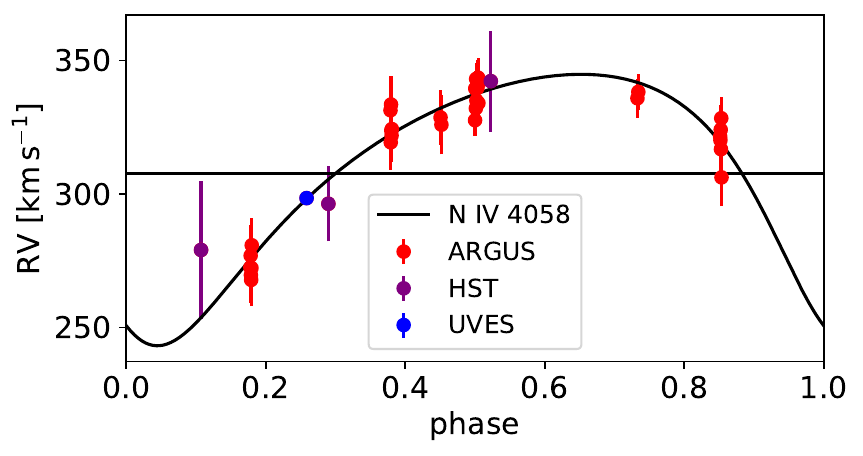}
\caption{As Fig.\,\ref{fig:OrbitTime}, but plotted as a function of phase and including the HST data and UVES datapoints.  } 
\label{fig:OrbitPhase}
\end{figure}

The lowest reduced $\chi^2$ (reduced $\chi^2 = 0.46$) is obtained for $P=17.20$\,d, which is refined to $P=17.2051\,$d during the minimisation. We obtain: $T_0 = 54737.8 \pm 1.5$\,[MJD], $V_0 = 307.6\pm2.8$\,\kms, $K_1 = 51\pm9\,$\kms, $\omega = 148.5 \pm 3.7^\circ$, $e = 0.31 \pm 0.08$. Figures\,\ref{fig:OrbitTime} and \ref{fig:OrbitPhase} compare this solution to the measurements. However, that acceptable solutions are found for all the periods listed above. We tried combining these RVs with those published by \citet{Schnurr2009} from IR data, but the period remains poorly constrained. Since the analysis involved a different spectral line than \NIV, we refrain from including the RVs from \citet{Schnurr2009} in our final analysis. Moreover, the preliminary 8.2\,d period derived by \citet{Schnurr2009} is not supported by our analysis. 

\begin{figure}
\centering
\includegraphics[width=.5\textwidth]{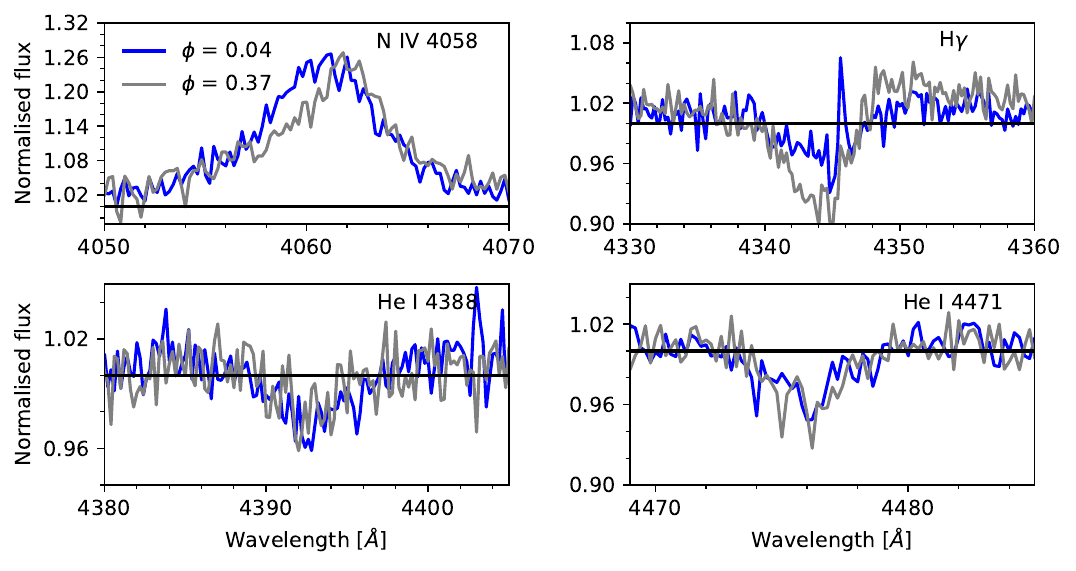}
\caption{Comparison of ARGUS spectra of R\,136~c taken during RV extremes, illustrating the apparent motion of the \NIV~line and the presence of seemingly static He\,{\sc i} absorption lines. } 
\label{fig:ExtSpecR136c}
\end{figure}

Another interesting fact is the presence of He\,{\sc i} absorption lines in the spectrum of R\,136~c (Fig.\,\ref{fig:ExtSpecR136c}). These He\,{\sc i} lines are seemingly static. The standard deviation of their RVs is 9.3\,\kms, comparable with the mean of the statistical error (7.7\,\kms). If these lines originate in the physical companion of the WR star in R\,136~c, then it must be a few times more massive than the WR star to avoid observed RV variability. This is somewhat in tension with the spectral type of the object, which is suggestive of a late-type O star or an early type B star. More likely, these lines belong to a distant tertiary source.  \citet{HB2012} noted that the ARGUS spaxel (which covers 0.52'', Sect.\,\ref{sec:data}) included two sources, with R\,136~c being significantly brighter. It is well possible that the second star produces the He\,{\sc i} absorption lines. 

The H$\gamma$ line shown in Fig.\,\ref{fig:ExtSpecR136c} likely stems from both the WN5h component and the late OB-type component. However, the variability seen in Balmer lines such as H$\gamma$ is likely dominated by the WNh5 component and its motion. WN5h stars, including a1, a2, and a3, typically show a combined emission+absorption profile in H$\beta$, H$\gamma$, and H$\delta$ \citep{Crowther2010}. Additional variability could stem from wind-wind collisions \citep[e.g.][]{Hill2000}, although this remains speculative without knowledge of the nature of the companion of the WN5h component. 

The nature of the secondary in R\,136~c thus remains unclear, and, in light of the discrepant RVs among the spectral lines (Fig.\,\ref{fig:DeltaRV}) and the multiple possible periods (Fig.\,\ref{fig:LSPeriod}), more data will be necessary to unambiguously derived the orbit. The fact that it is X-ray bright \citep{PZ2002} implies that the secondary is either another star with a strong wind (presumably an Of star or a WR star), or a compact object.

\section{Discussion}
\label{sec:discussion}

We can use the RVs measured in Sect.\,\ref{sec:analysis} to place constraints on possible companions to a1, a2, a3, and c. We use the RVs obtained for the \NIV~line, which has the smallest measurement errors after the \HeII~line, for which cross-contamination between a1 and a2 cannot be ruled out.
We perform Monte Carlo simulations to estimate the likelihood of specific binary configurations in reproducing the observed peak-to-peak RV variability. Specifically, for each pair of period $P$ and companion mass $M_2$ in the range $0.3 \le \log P[\,{\rm d}] \le 4.5$ and $2 \le M_2\,[M_\odot] \le 150$, respectively, we draw 1000 binaries from the following distributions: the eccentricities are drawn from a Gaussian distribution with a mean of 0.3 and a standard deviation of 0.2. In Appendix\,\ref{appendix:ecc}, we explore the impact of highly eccentric binaries. The primary mass is drawn from a uniform distribution in the range $100-300\,M_\odot$, the inclination $i$ is drawn uniformly on $\cos i$ (corresponding to a random orientation of the orbital plane), the argument of periastron $\omega$ is drawn uniformly in the range $0 - 2\,\pi$, and the time of periastron $T_0$ is drawn uniformly in the interval $0 - P$. The mass range $100-300\,M_\odot$ for the primaries is justified by WNh classification of our targets, which typically corresponds to $M \gtrsim 100\,M_\odot$, as well as by their previous mass determinations.  For each star and for each mock binary, the RVs are computed using the actual dates of observations for that star. For each mock binary, the errors on the set of RVs are assumed to be identical to the actual measured errors, and the mock RVs are modified assuming these errors. Like this, we form a series of 1000 peak-to-peak measurements corresponding to Eqs.\,(\ref{eq:p2p}) and (\ref{eq:Dpp}) for each $P, M_2$ pair.

\begin{figure}
\centering
\includegraphics[width=.5\textwidth]{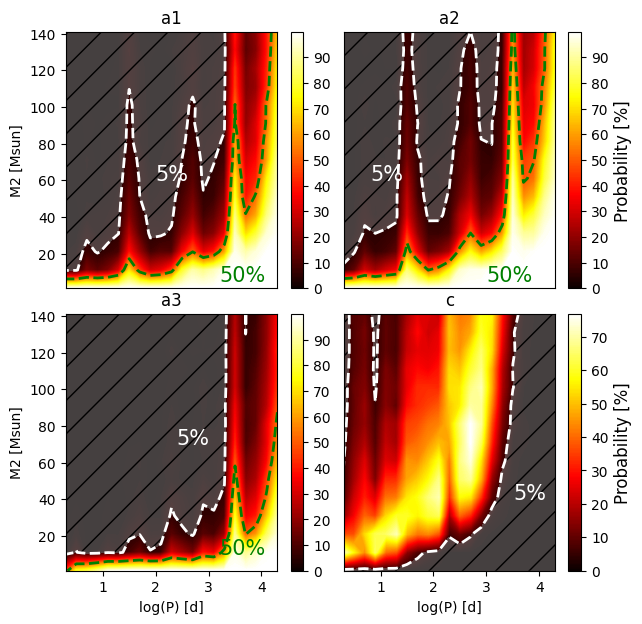}
\caption{Allowed configurations for companions for a1, a2, a3, and c using the RV measurements for the  \NIV~line. The panels for a1, a2, and a3, which were not classified as binaries here, show the probability of a binary with the given $M_2, P$ to produce values of $\Delta {\rm RV}_{{\rm pp}, \sigma}$ and $\Delta {\rm RV}_{{\rm pp}}$ below our observed values. The 5\% and 50\% thresholds are marked. For star c, which is classified as binary here,  shown is the probability that a binary in the given configuration would reproduce the observed $\Delta {\rm RV}_{\rm pp}$ within $\pm \sqrt{\sigma_i^2 + \sigma_j^2}$. Only 5\% thresholds are marked for clarity  (see text for details). } 
\label{fig:prob}
\end{figure}

\begin{figure}
\centering
\includegraphics[width=.5\textwidth]{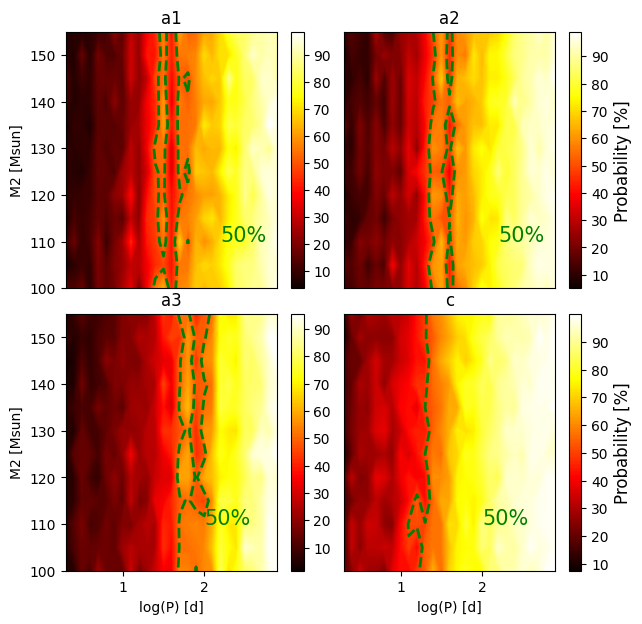}
\caption{Allowed configurations for companions for a1, a2, a3, and c for the case of two WR stars with identical spectra. Shown are the probabilities that the ratios between the largest and smallest FWHM values for the \NIV~line  are less than the observed values (Fig.\,\ref{fig:FWHM}). The 5\% and 50\% thresholds are marked. The probabilities drop sharply in comparison with the RV case (Fig.\,\ref{fig:prob} ). See text for details.   } 
\label{fig:probFWHM}
\end{figure}

Since stars a1, a2, and a3 are not flagged as binaries in our study, for these stars, Fig.\,\ref{fig:prob} shows the probability that a binary of a given $P, M_2$ would yield values of $\Delta {\rm RV}_{{\rm pp}, \sigma}$ and $\Delta {\rm RV}_{{\rm pp}}$ that are lower than those observed here for the \NIV~line. The shaded areas correspond to  configurations that can be rejected at 95\% probability, corresponding to the probability of such binaries to produce peak-to-peak RV variations larger than those observed.  The probabilities for star c, which was flagged as binary in our study, are discussed below.  Evidently, companions with masses $M_2 \lesssim 10\,M_\odot$ (corresponding to mass ratios $\lesssim 0.1$) cannot be ruled out at arbitrarily short periods. However, since our main focus is companions that could contribute significantly to the flux and bias the original mass estimates of these components, it is fair to focus our attention to $M_2 \gtrsim 50\,M_\odot$. For such stars, we can confidently rule out companions up to periods of a few years (separations $a \lesssim 10\,$au) at 95\% confidence, and up to tens of years ($a \lesssim 100\,$au) at 50\% confidence, unless the periods coincide with one of the aliases of the limited HST time series. Constraints for a3 are somewhat less stringent.

\citet{Kalari2022} obtained Speckle imaging of the R\,136 cluster and identified a companion at a projected separation of 2000\,au to a1 and a3, the former of which also identified by \citet{Lattanzi1994}, \citet{Hunter1995}, and \citet{Khorrami2017}. From figure 1 in \citet{Kalari2022}, we expect their study to be sensitive down to $\approx 1000\,$au. Hence, in conjunction with \citet{Kalari2022}, we cannot exclude massive companions in the range $10 \lesssim a\,[{\rm au}] \lesssim 1000$.

For star c, Fig.\,\ref{fig:prob} shows the probability as a function of $P, M_2$ that a star would reproduce the observed peak-to-peak variability. Specifically, we require that $\left| \Delta {\rm RV}_{\rm pp, mock} - \Delta {\rm RV}_{\rm pp, obs} \right| < \sqrt{\sigma_i^2 + \sigma_j^2}$, where $\sigma_i, \sigma_j$ are the errors on the RVs which produce $\Delta {\rm RV}_{\rm pp, obs}$. This limits, within 95\% confidence, the range of acceptable companion masses and periods to R\,136~c. The results are consistent with the period of 17.20\,d  derived in our study.

Finally, we consider the case of two WR-like stars with similar line profiles and light contributions, and consider the FWHM variability that could be expected in this case (see discussion in Sect.\,\ref{sec:analysis}). We focus on the \NIV~line, to prevent possible cross-contamination between a1 and a2 in the \HeII~line impacting our results.  Like the exercise above, for each pair of $M_2, P$, we draw 1000 binaries following the same distributions as before, focusing on $100 \le M_2/M_\odot \le 150$ and $0.3 \le \log P\,[{\rm d}] \le 3$. We fit the underlying \NIV~profile of a1, a2, a3, and c with Gaussians. Then, we compute the FWHM of a Gaussian comprising of the sum of two such Gaussians that are shifted relative to one another by $\Delta {\rm RV}$. We fit a quadratic function to FWHM($\Delta {\rm RV}$) to obtain an analytical relation between the FWHM and $\Delta {\rm RV}_{\rm pp}$ for each star. For each $\Delta {\rm RV}_{\rm pp}$ value in a given simulation of a $M_2, P$ pair, we classify the mock binary as binary if the ratio of the FWHM of the \NIV~line to that of the unshifted Gaussian exceeds the values we observe. The results are shown in Fig.\,\ref{fig:probFWHM}.

Evidently, the probability to detect companions with similar spectra sharply drops in comparison with binaries containing only one WR star. Only very short period ($\lesssim 3\,$d) can be rejected at high probability (90\%); the 50\% thresholds lie at periods of the order of 10-100\,d. The results are insensitive to $M_2$, recalling that the underlying assumption here is that the two stars have similar spectra. If the secondary is significantly fainter or does not show \NIV~or \HeII~in emission, then the RVs become a sensitive probe, and Fig.\,\ref{fig:prob} becomes the relevant diagnostic. We conclude that close companions of a similar spectral type cannot be readily excluded from the current data.  A longer time coverage and higher data quality should allow for more stringent constraints in this case.

The absence of strong X-ray emission in a1, a2, and a3 does not favour the presence of close massive companions. For example, the X-ray luminosity of the WN5+WN5 binary Mk\,34, with orbital period of 155\,d,  is $\approx 10^{35}\,{\rm erg}\,{\rm s}^{-1}$ \citep{Pollock2018, Tehrani2019}, which exceeds the combined X-ray luminosity of a1, a2, and nearby targets by an order of magnitude \citep{Crowther2022}. However, the colliding wind phenomenon occurs only in a subset of massive binaries -- the X-ray emission of  a majority of known spectroscopic binaries does not exceed average value $L_{\rm X} / L_{\rm bol} \sim 10^{-7}$ \citep{Osk2005, Sana2006, Rauw2016, Nebot2018, Crowther2022}. For example, R\,144 is a colliding-wind binary hosting two WR stars bound on a 74\,d period which does not exhibit strong X-ray excess \citep{Shenar2021}. \citet{Naze2009} noted that, for the short period massive binaries ($P_{\rm orb} \lesssim 30$\,d), the prevalence of enhanced X-rays is lower compared to longer period binaries.  The physical reason could be the braking of stellar winds by the radiation of companions in close binaries which dramatically reduces the strength of the wind collision \citep{Gayley1997}, or that the collision occurs withing the wind acceleration zone \citep{Sana2004}. Furthermore, \citet{Krticka2015} suggest that intrinsic X-ray emission could lead to wind inhibition in massive binaries. Hence, generally, while X-ray excess provides indirect support for a companion with a powerful wind, lack of X-ray excess does not suffice to reject such a companion.

\section{Summary}
\label{sec:summary}

We investigated whether some of the most massive stars reported to date -- R\,136~a1, a2, a3, and c -- may be binaries that host two massive stars, which would affect their previous mass determinations. To this end, we collected three epochs of optical spectroscopy over 1.5\,yr in the years 2020-2021 with the STIS instrument on the HST to search for RV and EW variations or other indications for binary motion. These data were combined with an additional epoch in 2012 acquired by \citet{Crowther2016} to form a 10-year baseline for our study. For R\,136~c, we combined these data with archival FLAMES and UVES data to derive a preliminary orbital solution.

The data are not readily suggestive of close companions to the stars a1, a2, or a3. We can rule out companions more massive than $\approx 50\,M_\odot$ out to orbital periods $\lesssim 1-3$ years ($a \lesssim 5-10\,$au) at 95\% confidence, or periods of tens of years ($a \lesssim 100\,$au) at 50\% confidence. In conjunction with previous imaging studies \citep{Khorrami2017, Kalari2022}, additional companions could only reside in the range $\approx 10 - 1000\,$au. However, "twin companions" with similar light contributions and spectral appearance could avoid detection down to much shorter periods ($P \gtrsim 10\,$d), though no direct indications for such companions (e.g., from the spectral appearance or X-ray behaviour) are noted. Hence, the masses of a1, a2, and a3 may still be considered to be $\gtrsim 150\,M_\odot$ \citep{Crowther2010, Bestenlehner2020, Brands2022}, although the faint visual companions to a1 and a3 \citep{Hunter1995, Kalari2022}  may lead to a modest downward revision of the masses.

In contrast, R\,136~c is classified as a binary in our study. This is consistent with previous indications for binarity reported by \citet{Schnurr2009} and \citet{HB2012} , who reported this object to be a binary candidate based on IR and VIS spectroscopy. Combining archival data with the new one, we propose a tentative period of 17.2\,d for R\,136~c, though more data will be needed to robustly constrain the orbital configuration and the true nature of the companion.  Given the X-ray brightness of the system and the rarity of such very massive binaries, future monitoring of R\,136~c would  yield important constraints on the masses of the most massive stars and on massive binary evolution.

\begin{acknowledgements}
We thank the anonymous referee for improving our manuscript.
TS acknowledges support from the European Union's Horizon 2020 under the Marie Sk\l{}odowska-Curie grant agreement No 101024605 and from the Comunidad de Madrid (2022-T1/TIC-24117).
This publication was made possible through the support of an LSSTC Catalyst Fellowship to K.A.B., funded through Grant 62192 from the John Templeton Foundation to LSST Corporation. The opinions expressed in this publication are those of the authors and do not necessarily reflect the views of LSSTC or the John Templeton Foundation. AACS is supported by the Deutsche Forschungsgemeinschaft (DFG, German Research Foundation) in the form of an Emmy Noether Research Group -- Project-ID 445674056 (SA4064/1-1, PI Sander) and acknowledges further support from the Federal Ministry of Education and Research (BMBF) and
the Baden-W\"urttemberg Ministry of Science as part of the Excellence Strategy of the German Federal and State Governments. PAC is supported by the Science and Technology Facilities Council research grant ST/V000853/1 (PI. V. Dhillon).  F.N. acknowledges grant PID2019-105552RB-C4 funded by the Spanish  MCIN/AEI/ 10.13039/501100011033.
This publication was partially supported by the International Space Science Institute (ISSI) in Bern, through ISSI International Team project 512 (Multiwavelength View on Massive Stars in the Era of Multimessenger Astronomy (PI Oskinova). We appreciate discussions with Andy Pollock on the potential period of R136c from X-ray observations, and with Jes\'{u}s Ma\'{i}z Apell\'{a}niz and Danny Lennon on reduction methods of the STIS dataset.
\end{acknowledgements}

\bibliographystyle{aa}
\bibliography{papers}

\begin{thebibliography}{92}
\expandafter\ifx\csname natexlab\endcsname\relax\def\natexlab#1{#1}\fi

\bibitem[{{Barb{\'a}} {et~al.}(2022){Barb{\'a}}, {Gamen}, {Mart{\'\i}n-Ravelo},
  {Arias}, \& {Morrell}}]{Barba2022}
{Barb{\'a}}, R.~H., {Gamen}, R.~C., {Mart{\'\i}n-Ravelo}, P., {Arias}, J.~I.,
  \& {Morrell}, N.~I. 2022, \mnras, 516, 1149

\bibitem[{{Bastian} \& {Lardo}(2018)}]{Bastian2018}
{Bastian}, N. \& {Lardo}, C. 2018, \araa, 56, 83

\bibitem[{{Bestenlehner} {et~al.}(2022){Bestenlehner}, {Crowther}, {Broos},
  {Pollock}, \& {Townsley}}]{Bestenlehner2022}
{Bestenlehner}, J.~M., {Crowther}, P.~A., {Broos}, P.~S., {Pollock}, A. M.~T.,
  \& {Townsley}, L.~K. 2022, \mnras, 510, 6133

\bibitem[{{Bestenlehner} {et~al.}(2020){Bestenlehner}, {Crowther},
  {Caballero-Nieves}, {Schneider}, {Sim{\'o}n-D{\'\i}az}, {Brands}, {de Koter},
  {Gr{\"a}fener}, {Herrero}, {Langer}, {Lennon}, {Ma{\'\i}z Apell{\'a}niz},
  {Puls}, \& {Vink}}]{Bestenlehner2020}
{Bestenlehner}, J.~M., {Crowther}, P.~A., {Caballero-Nieves}, S.~M., {et~al.}
  2020, \mnras, 499, 1918

\bibitem[{{Bestenlehner} {et~al.}(2014){Bestenlehner}, {Gr{\"a}fener}, {Vink},
  {Najarro}, {de Koter}, {Sana}, {Evans}, {Crowther}, {H{\'e}nault-Brunet},
  {Herrero}, {Langer}, {Schneider}, {Sim{\'o}n-D{\'\i}az}, {Taylor}, \&
  {Walborn}}]{Bestenlehner2014}
{Bestenlehner}, J.~M., {Gr{\"a}fener}, G., {Vink}, J.~S., {et~al.} 2014, \aap,
  570, A38

\bibitem[{{Bonanos} {et~al.}(2004){Bonanos}, {Stanek}, {Udalski},
  {Wyrzykowski}, {{\.Z}ebru{\'n}}, {Kubiak}, {Szyma{\'n}ski}, {Szewczyk},
  {Pietrzy{\'n}ski}, \& {Soszy{\'n}ski}}]{Bonanos2004}
{Bonanos}, A.~Z., {Stanek}, K.~Z., {Udalski}, A., {et~al.} 2004, \apjl, 611,
  L33

\bibitem[{{Brands} {et~al.}(2022){Brands}, {de Koter}, {Bestenlehner},
  {Crowther}, {Sundqvist}, {Puls}, {Caballero-Nieves}, {Abdul-Masih},
  {Driessen}, {Garc{\'\i}a}, {Geen}, {Gr{\"a}fener}, {Hawcroft}, {Kaper},
  {Keszthelyi}, {Langer}, {Sana}, {Schneider}, {Shenar}, \&
  {Vink}}]{Brands2022}
{Brands}, S.~A., {de Koter}, A., {Bestenlehner}, J.~M., {et~al.} 2022, \aap,
  663, A36

\bibitem[{{Brown} {et~al.}(1978){Brown}, {McLean}, \& {Emslie}}]{Brown1978}
{Brown}, J.~C., {McLean}, I.~S., \& {Emslie}, A.~G. 1978, \aap, 68, 415

\bibitem[{{Cassinelli} {et~al.}(1981){Cassinelli}, {Mathis}, \&
  {Savage}}]{Cassinelli1981}
{Cassinelli}, J.~P., {Mathis}, J.~S., \& {Savage}, B.~D. 1981, Science, 212,
  1497

\bibitem[{{Chalabaev} \& {Maillard}(1983)}]{Chalabaev1983}
{Chalabaev}, A. \& {Maillard}, J.~P. 1983, \aap, 127, 279

\bibitem[{{Cox} {et~al.}(2005){Cox}, {Kaper}, {Foing}, \&
  {Ehrenfreund}}]{Cox2005}
{Cox}, N.~L.~J., {Kaper}, L., {Foing}, B.~H., \& {Ehrenfreund}, P. 2005, \aap,
  438, 187

\bibitem[{{Crowther} {et~al.}(2022){Crowther}, {Broos}, {Townsley}, {Pollock},
  {Tehrani}, \& {Gagn{\'e}}}]{Crowther2022}
{Crowther}, P.~A., {Broos}, P.~S., {Townsley}, L.~K., {et~al.} 2022, \mnras,
  515, 4130

\bibitem[{{Crowther} {et~al.}(2016){Crowther}, {Caballero-Nieves}, {Bostroem},
  {Ma{\'\i}z Apell{\'a}niz}, {Schneider}, {Walborn}, {Angus}, {Brott},
  {Bonanos}, {de Koter}, {de Mink}, {Evans}, {Gr{\"a}fener}, {Herrero},
  {Howarth}, {Langer}, {Lennon}, {Puls}, {Sana}, \& {Vink}}]{Crowther2016}
{Crowther}, P.~A., {Caballero-Nieves}, S.~M., {Bostroem}, K.~A., {et~al.} 2016,
  \mnras, 458, 624

\bibitem[{{Crowther} \& {Dessart}(1998)}]{Crowther1998}
{Crowther}, P.~A. \& {Dessart}, L. 1998, \mnras, 296, 622

\bibitem[{{Crowther} {et~al.}(2010){Crowther}, {Schnurr}, {Hirschi}, {Yusof},
  {Parker}, {Goodwin}, \& {Kassim}}]{Crowther2010}
{Crowther}, P.~A., {Schnurr}, O., {Hirschi}, R., {et~al.} 2010, \mnras, 408,
  731

\bibitem[{{Crowther} \& {Walborn}(2011)}]{Crowther2011}
{Crowther}, P.~A. \& {Walborn}, N.~R. 2011, \mnras, 416, 1311

\bibitem[{{de Koter} {et~al.}(1997){de Koter}, {Heap}, \&
  {Hubeny}}]{deKoter1997}
{de Koter}, A., {Heap}, S.~R., \& {Hubeny}, I. 1997, \apj, 477, 792

\bibitem[{{Doran} {et~al.}(2013){Doran}, {Crowther}, {de Koter}, {Evans},
  {McEvoy}, {Walborn}, {Bastian}, {Bestenlehner}, {Gr{\"a}fener}, {Herrero},
  {K{\"o}hler}, {Ma{\'\i}z Apell{\'a}niz}, {Najarro}, {Puls}, {Sana},
  {Schneider}, {Taylor}, {van Loon}, \& {Vink}}]{Doran2013}
{Doran}, E.~I., {Crowther}, P.~A., {de Koter}, A., {et~al.} 2013, \aap, 558,
  A134

\bibitem[{{Dsilva} {et~al.}(2020){Dsilva}, {Shenar}, {Sana}, \&
  {Marchant}}]{Dsilva2020}
{Dsilva}, K., {Shenar}, T., {Sana}, H., \& {Marchant}, P. 2020, \aap, 641, A26

\bibitem[{{Dsilva} {et~al.}(2022){Dsilva}, {Shenar}, {Sana}, \&
  {Marchant}}]{Dsilva2022}
{Dsilva}, K., {Shenar}, T., {Sana}, H., \& {Marchant}, P. 2022, \aap, 664, A93

\bibitem[{{Dsilva} {et~al.}(2023){Dsilva}, {Shenar}, {Sana}, \&
  {Marchant}}]{Dsilva2023}
{Dsilva}, K., {Shenar}, T., {Sana}, H., \& {Marchant}, P. 2023, \aap, 674, A88

\bibitem[{{Evans} {et~al.}(2011){Evans}, {Taylor}, {H{\'e}nault-Brunet},
  {Sana}, {de Koter}, {Sim{\'o}n-D{\'\i}az}, {Carraro}, {Bagnoli}, {Bastian},
  {Bestenlehner}, {Bonanos}, {Bressert}, {Brott}, {Campbell}, {Cantiello},
  {Clark}, {Costa}, {Crowther}, {de Mink}, {Doran}, {Dufton}, {Dunstall},
  {Friedrich}, {Garcia}, {Gieles}, {Gr{\"a}fener}, {Herrero}, {Howarth},
  {Izzard}, {Langer}, {Lennon}, {Ma{\'\i}z Apell{\'a}niz}, {Markova},
  {Najarro}, {Puls}, {Ramirez}, {Sab{\'\i}n-Sanjuli{\'a}n}, {Smartt}, {Stroud},
  {van Loon}, {Vink}, \& {Walborn}}]{Evans2011}
{Evans}, C.~J., {Taylor}, W.~D., {H{\'e}nault-Brunet}, V., {et~al.} 2011, \aap,
  530, A108

\bibitem[{{Figer}(2005)}]{Figer2005}
{Figer}, D.~F. 2005, \nat, 434, 192

\bibitem[{{Figer} {et~al.}(2002){Figer}, {Najarro}, {Gilmore}, {Morris}, {Kim},
  {Serabyn}, {McLean}, {Gilbert}, {Graham}, {Larkin}, {Levenson}, \&
  {Teplitz}}]{Figer2002}
{Figer}, D.~F., {Najarro}, F., {Gilmore}, D., {et~al.} 2002, \apj, 581, 258

\bibitem[{{Fryer} {et~al.}(2001){Fryer}, {Woosley}, \& {Heger}}]{Fryer2001}
{Fryer}, C.~L., {Woosley}, S.~E., \& {Heger}, A. 2001, \apj, 550, 372

\bibitem[{{Gayley} {et~al.}(1997){Gayley}, {Owocki}, \& {Cranmer}}]{Gayley1997}
{Gayley}, K.~G., {Owocki}, S.~P., \& {Cranmer}, S.~R. 1997, \apj, 475, 786

\bibitem[{{Gieles} {et~al.}(2018){Gieles}, {Charbonnel}, {Krause},
  {H{\'e}nault-Brunet}, {Agertz}, {Lamers}, {Bastian}, {Gualandris}, {Zocchi},
  \& {Petts}}]{Gieles2018}
{Gieles}, M., {Charbonnel}, C., {Krause}, M. G.~H., {et~al.} 2018, \mnras, 478,
  2461

\bibitem[{{Guerrero} \& {Chu}(2008)}]{Guerrero2008}
{Guerrero}, M.~A. \& {Chu}, Y.-H. 2008, \apjs, 177, 216

\bibitem[{{Hainich} {et~al.}(2014){Hainich}, {R{\"u}hling}, {Todt}, {Oskinova},
  {Liermann}, {Gr{\"a}fener}, {Foellmi}, {Schnurr}, \& {Hamann}}]{Hainich2014}
{Hainich}, R., {R{\"u}hling}, U., {Todt}, H., {et~al.} 2014, \aap, 565, A27

\bibitem[{{Hamann} \& {Gr{\"a}fener}(2003)}]{Hamann2003}
{Hamann}, W.~R. \& {Gr{\"a}fener}, G. 2003, \aap, 410, 993

\bibitem[{{Heap} {et~al.}(1994){Heap}, {Ebbets}, {Malumuth}, {Maran}, {de
  Koter}, \& {Hubeny}}]{Heap1994}
{Heap}, S.~R., {Ebbets}, D., {Malumuth}, E.~M., {et~al.} 1994, \apjl, 435, L39

\bibitem[{{H{\'e}nault-Brunet} {et~al.}(2012){H{\'e}nault-Brunet}, {Evans},
  {Sana}, {Gieles}, {Bastian}, {Ma{\'\i}z Apell{\'a}niz}, {Markova}, {Taylor},
  {Bressert}, {Crowther}, \& {van Loon}}]{HB2012}
{H{\'e}nault-Brunet}, V., {Evans}, C.~J., {Sana}, H., {et~al.} 2012, \aap, 546,
  A73

\bibitem[{{Hill} {et~al.}(2000){Hill}, {Moffat}, {St-Louis}, \&
  {Bartzakos}}]{Hill2000}
{Hill}, G.~M., {Moffat}, A.~F.~J., {St-Louis}, N., \& {Bartzakos}, P. 2000,
  \mnras, 318, 402

\bibitem[{{Hunter} {et~al.}(1995){Hunter}, {Shaya}, {Holtzman}, {Light},
  {O'Neil}, \& {Lynds}}]{Hunter1995}
{Hunter}, D.~A., {Shaya}, E.~J., {Holtzman}, J.~A., {et~al.} 1995, \apj, 448,
  179

\bibitem[{{Kalari} {et~al.}(2022){Kalari}, {Horch}, {Salinas}, {Vink},
  {Andersen}, {Bestenlehner}, \& {Rubio}}]{Kalari2022}
{Kalari}, V.~M., {Horch}, E.~P., {Salinas}, R., {et~al.} 2022, \apj, 935, 162

\bibitem[{{Khorrami} {et~al.}(2017){Khorrami}, {Vakili}, {Lanz}, {Langlois},
  {Lagadec}, {Meyer}, {Robbe-Dubois}, {Abe}, {Avenhaus}, {Beuzit}, {Gratton},
  {Mouillet}, {Orign{\'e}}, {Petit}, \& {Ramos}}]{Khorrami2017}
{Khorrami}, Z., {Vakili}, F., {Lanz}, T., {et~al.} 2017, \aap, 602, A56

\bibitem[{{Knigge} {et~al.}(2008){Knigge}, {Dieball}, {Ma{\'\i}z
  Apell{\'a}niz}, {Long}, {Zurek}, \& {Shara}}]{Knigge2008}
{Knigge}, C., {Dieball}, A., {Ma{\'\i}z Apell{\'a}niz}, J., {et~al.} 2008, in
  Dynamical Evolution of Dense Stellar Systems, ed. E.~{Vesperini},
  M.~{Giersz}, \& A.~{Sills}, Vol. 246, 321--325

\bibitem[{{Koenigsberger} {et~al.}(2014){Koenigsberger}, {Morrell}, {Hillier},
  {Gamen}, {Schneider}, {Gonz{\'a}lez-Jim{\'e}nez}, {Langer}, \&
  {Barb{\'a}}}]{Koenigsberger2014}
{Koenigsberger}, G., {Morrell}, N., {Hillier}, D.~J., {et~al.} 2014, \aj, 148,
  62

\bibitem[{{Krti{\v{c}}ka} {et~al.}(2015){Krti{\v{c}}ka}, {Kub{\'a}t}, \&
  {Krti{\v{c}}kov{\'a}}}]{Krticka2015}
{Krti{\v{c}}ka}, J., {Kub{\'a}t}, J., \& {Krti{\v{c}}kov{\'a}}, I. 2015, \aap,
  579, A111

\bibitem[{{Lamontagne} {et~al.}(1996){Lamontagne}, {Moffat}, {Drissen},
  {Robert}, \& {Matthews}}]{Lamontagne1996}
{Lamontagne}, R., {Moffat}, A. F.~J., {Drissen}, L., {Robert}, C., \&
  {Matthews}, J.~M. 1996, \aj, 112, 2227

\bibitem[{{Langer}(2012)}]{Langer2012}
{Langer}, N. 2012, \araa, 50, 107

\bibitem[{{Larson} \& {Starrfield}(1971)}]{Larson1971}
{Larson}, R.~B. \& {Starrfield}, S. 1971, \aap, 13, 190

\bibitem[{{Lattanzi} {et~al.}(1994){Lattanzi}, {Hershey}, {Burg}, {Taff},
  {Holfeltz}, {Bucciarelli}, {Evans}, {Gilmozzi}, {Pringle}, \&
  {Walborn}}]{Lattanzi1994}
{Lattanzi}, M.~G., {Hershey}, J.~L., {Burg}, R., {et~al.} 1994, \apjl, 427, L21

\bibitem[{{Lennon} {et~al.}(2021){Lennon}, {Ma{\'\i}z Apell{\'a}niz},
  {Irrgang}, {Bohlin}, {Deustua}, {Dufton}, {Sim{\'o}n-D{\'\i}az}, {Herrero},
  {Casares}, {Mu{\~n}oz-Darias}, {Smartt}, {Gonz{\'a}lez Hern{\'a}ndez}, \& {de
  Burgos}}]{Lennon2021}
{Lennon}, D.~J., {Ma{\'\i}z Apell{\'a}niz}, J., {Irrgang}, A., {et~al.} 2021,
  \aap, 649, A167

\bibitem[{{L{\'e}pine} \& {Moffat}(1999)}]{Lepine1999}
{L{\'e}pine}, S. \& {Moffat}, A. F.~J. 1999, \apj, 514, 909

\bibitem[{{Lohr} {et~al.}(2018){Lohr}, {Clark}, {Najarro}, {Patrick},
  {Crowther}, \& {Evans}}]{Lohr2018}
{Lohr}, M.~E., {Clark}, J.~S., {Najarro}, F., {et~al.} 2018, \aap, 617, A66

\bibitem[{{Luehrs}(1997)}]{Luehrs1997}
{Luehrs}, S. 1997, \pasp, 109, 504

\bibitem[{{Madura} {et~al.}(2012){Madura}, {Gull}, {Owocki}, {Groh}, {Okazaki},
  \& {Russell}}]{Madura2012}
{Madura}, T.~I., {Gull}, T.~R., {Owocki}, S.~P., {et~al.} 2012, \mnras, 420,
  2064

\bibitem[{{Mahy} {et~al.}(2020){Mahy}, {Sana}, {Abdul-Masih}, {Almeida},
  {Langer}, {Shenar}, {de Koter}, {de Mink}, {de Wit}, {Grin}, {Evans},
  {Moffat}, {Schneider}, {Barb{\'a}}, {Clark}, {Crowther}, {Gr{\"a}fener},
  {Lennon}, {Tramper}, \& {Vink}}]{Mahy2020a}
{Mahy}, L., {Sana}, H., {Abdul-Masih}, M., {et~al.} 2020, \aap, 634, A118

\bibitem[{{Maiz-Apellaniz}(2005)}]{Maiz-Apellaniz2005}
{Maiz-Apellaniz}, J. 2005, {MULTISPEC: A Code for the Extraction of Slitless
  Spectra in Crowded Fields}, Instrument Science Report STIS 2005-02, 18 pages

\bibitem[{{Martins} {et~al.}(2008){Martins}, {Hillier}, {Paumard},
  {Eisenhauer}, {Ott}, \& {Genzel}}]{Martins2008}
{Martins}, F., {Hillier}, D.~J., {Paumard}, T., {et~al.} 2008, \aap, 478, 219

\bibitem[{{Massey} \& {Hunter}(1998)}]{Massey1998}
{Massey}, P. \& {Hunter}, D.~A. 1998, \apj, 493, 180

\bibitem[{{Moffat} \& {Bobert}(1992)}]{Moffat1992}
{Moffat}, A.~F.~J. \& {Bobert}, C. 1992, in Astronomical Society of the Pacific
  Conference Series, Vol.~22, Nonisotropic and Variable Outflows from Stars,
  ed. L.~{Drissen}, C.~{Leitherer}, \& A.~{Nota}, 203

\bibitem[{{Najarro} {et~al.}(2004){Najarro}, {Figer}, {Hillier}, \&
  {Kudritzki}}]{Najarro2004}
{Najarro}, F., {Figer}, D.~F., {Hillier}, D.~J., \& {Kudritzki}, R.~P. 2004,
  \apjl, 611, L105

\bibitem[{{Naz{\'e}}(2009)}]{Naze2009}
{Naz{\'e}}, Y. 2009, \aap, 506, 1055

\bibitem[{{Nebot G{\'o}mez-Mor{\'a}n} \& {Oskinova}(2018)}]{Nebot2018}
{Nebot G{\'o}mez-Mor{\'a}n}, A. \& {Oskinova}, L.~M. 2018, \aap, 620, A89

\bibitem[{{Oey} \& {Clarke}(2005)}]{Oey2005}
{Oey}, M.~S. \& {Clarke}, C.~J. 2005, \apjl, 620, L43

\bibitem[{{Oskinova}(2005)}]{Osk2005}
{Oskinova}, L.~M. 2005, \mnras, 361, 679

\bibitem[{{Pollock} {et~al.}(2018){Pollock}, {Crowther}, {Tehrani}, {Broos}, \&
  {Townsley}}]{Pollock2018}
{Pollock}, A.~M.~T., {Crowther}, P.~A., {Tehrani}, K., {Broos}, P.~S., \&
  {Townsley}, L.~K. 2018, \mnras, 474, 3228

\bibitem[{{Portegies Zwart} {et~al.}(2002){Portegies Zwart}, {Pooley}, \&
  {Lewin}}]{PZ2002}
{Portegies Zwart}, S.~F., {Pooley}, D., \& {Lewin}, W. H.~G. 2002, \apj, 574,
  762

\bibitem[{{Quimby} {et~al.}(2011){Quimby}, {Kulkarni}, {Kasliwal}, {Gal-Yam},
  {Arcavi}, {Sullivan}, {Nugent}, {Thomas}, {Howell}, {Nakar}, {Bildsten},
  {Theissen}, {Law}, {Dekany}, {Rahmer}, {Hale}, {Smith}, {Ofek}, {Zolkower},
  {Velur}, {Walters}, {Henning}, {Bui}, {McKenna}, {Poznanski}, {Cenko}, \&
  {Levitan}}]{Quimby2011}
{Quimby}, R.~M., {Kulkarni}, S.~R., {Kasliwal}, M.~M., {et~al.} 2011, \nat,
  474, 487

\bibitem[{{Ramachandran} {et~al.}(2019){Ramachandran}, {Hamann}, {Oskinova},
  {Gallagher}, {Hainich}, {Shenar}, {Sand er}, {Todt}, \&
  {Fulmer}}]{Ramachandran2019}
{Ramachandran}, V., {Hamann}, W.~R., {Oskinova}, L.~M., {et~al.} 2019, \aap,
  625, A104

\bibitem[{{Rauw} {et~al.}(2004){Rauw}, {De Becker}, {Naz{\'e}}, {Crowther},
  {Gosset}, {Sana}, {van der Hucht}, {Vreux}, \& {Williams}}]{Rauw2004}
{Rauw}, G., {De Becker}, M., {Naz{\'e}}, Y., {et~al.} 2004, \aap, 420, L9

\bibitem[{{Rauw} \& {Naz{\'e}}(2016)}]{Rauw2016}
{Rauw}, G. \& {Naz{\'e}}, Y. 2016, Advances in Space Research, 58, 761

\bibitem[{{Richardson} {et~al.}(2016){Richardson}, {Shenar}, {Roy-Loubier},
  {Schaefer}, {Moffat}, {St-Louis}, {Gies}, {Farrington}, {Hill}, {Williams},
  {Gordon}, {Pablo}, \& {Ramiaramanantsoa}}]{Richardson2016}
{Richardson}, N.~D., {Shenar}, T., {Roy-Loubier}, O., {et~al.} 2016, \mnras,
  461, 4115

\bibitem[{{Robert} {et~al.}(1992){Robert}, {Moffat}, {Drissen}, {Lamontagne},
  {Seggewiss}, {Niemela}, {Cerruti}, {Barrett}, {Bailey}, {Garcia}, \&
  {Tapia}}]{Robert1992}
{Robert}, C., {Moffat}, A. F.~J., {Drissen}, L., {et~al.} 1992, \apj, 397, 277

\bibitem[{{Rubio-D{\'\i}ez} {et~al.}(2017){Rubio-D{\'\i}ez}, {Najarro},
  {Garc{\'\i}a}, \& {Sundqvist}}]{RubioDiez2017}
{Rubio-D{\'\i}ez}, M.~M., {Najarro}, F., {Garc{\'\i}a}, M., \& {Sundqvist},
  J.~O. 2017, in The Lives and Death-Throes of Massive Stars, ed. J.~J.
  {Eldridge}, J.~C. {Bray}, L.~A.~S. {McClelland}, \& L.~{Xiao}, Vol. 329,
  131--135

\bibitem[{{Sana} {et~al.}(2013){Sana}, {de Koter}, {de Mink}, {Dunstall},
  {Evans}, {H{\'e}nault-Brunet}, {Ma{\'\i}z Apell{\'a}niz},
  {Ram{\'\i}rez-Agudelo}, {Taylor}, {Walborn}, {Clark}, {Crowther}, {Herrero},
  {Gieles}, {Langer}, {Lennon}, \& {Vink}}]{Sana2013b}
{Sana}, H., {de Koter}, A., {de Mink}, S.~E., {et~al.} 2013, \aap, 550, A107

\bibitem[{Sana {et~al.}(2012)Sana, de~Mink, de~Koter, Langer, Evans, Gieles,
  Gosset, Izzard, {Le Bouquin}, \& Schneider}]{Sana2012}
Sana, H., de~Mink, S.~E., de~Koter, A., {et~al.} 2012, Science, 337, 444

\bibitem[{{Sana} {et~al.}(2011){Sana}, {Le Bouquin}, {De Becker}, {Berger}, {de
  Koter}, \& {M{\'e}rand}}]{Sana2011}
{Sana}, H., {Le Bouquin}, J.~B., {De Becker}, M., {et~al.} 2011, \apjl, 740,
  L43

\bibitem[{{Sana} {et~al.}(2006){Sana}, {Rauw}, {Naz{\'e}}, {Gosset}, \&
  {Vreux}}]{Sana2006}
{Sana}, H., {Rauw}, G., {Naz{\'e}}, Y., {Gosset}, E., \& {Vreux}, J.~M. 2006,
  \mnras, 372, 661

\bibitem[{{Sana} {et~al.}(2004){Sana}, {Stevens}, {Gosset}, {Rauw}, \&
  {Vreux}}]{Sana2004}
{Sana}, H., {Stevens}, I.~R., {Gosset}, E., {Rauw}, G., \& {Vreux}, J.~M. 2004,
  \mnras, 350, 809

\bibitem[{{Sander} {et~al.}(2015){Sander}, {Shenar}, {Hainich},
  {G{\'\i}menez-Garc{\'\i}a}, {Todt}, \& {Hamann}}]{Sander2015}
{Sander}, A., {Shenar}, T., {Hainich}, R., {et~al.} 2015, \aap, 577, A13

\bibitem[{{Savage} {et~al.}(1983){Savage}, {Fitzpatrick}, {Cassinelli}, \&
  {Ebbets}}]{Savage1983}
{Savage}, B.~D., {Fitzpatrick}, E.~L., {Cassinelli}, J.~P., \& {Ebbets}, D.~C.
  1983, \apj, 273, 597

\bibitem[{{Schnurr} {et~al.}(2008){Schnurr}, {Casoli}, {Chen{\'e}}, {Moffat},
  \& {St-Louis}}]{Schnurr2008}
{Schnurr}, O., {Casoli}, J., {Chen{\'e}}, A.~N., {Moffat}, A.~F.~J., \&
  {St-Louis}, N. 2008, \mnras, 389, L38

\bibitem[{{Schnurr} {et~al.}(2009){Schnurr}, {Chen{\'e}}, {Casoli}, {Moffat},
  \& {St-Louis}}]{Schnurr2009}
{Schnurr}, O., {Chen{\'e}}, A.~N., {Casoli}, J., {Moffat}, A.~F.~J., \&
  {St-Louis}, N. 2009, \mnras, 397, 2049

\bibitem[{{Shenar} {et~al.}(2017{\natexlab{a}}){Shenar}, {Oskinova},
  {J{\"a}rvinen}, {Luckas}, {Hainich}, {Todt}, {Hubrig}, {Sand er}, {Ilyin}, \&
  {Hamann}}]{Shenar2017}
{Shenar}, T., {Oskinova}, L.~M., {J{\"a}rvinen}, S.~P., {et~al.}
  2017{\natexlab{a}}, \aap, 606, A91

\bibitem[{{Shenar} {et~al.}(2017{\natexlab{b}}){Shenar}, {Richardson},
  {Sablowski}, {Hainich}, {Sana}, {Moffat}, {Todt}, {Hamann}, {Oskinova},
  {Sander}, {Tramper}, {Langer}, {Bonanos}, {de Mink}, {Gr{\"a}fener},
  {Crowther}, {Vink}, {Almeida}, {de Koter}, {Barb{\'a}}, {Herrero}, \&
  {Ulaczyk}}]{Shenar2017b}
{Shenar}, T., {Richardson}, N.~D., {Sablowski}, D.~P., {et~al.}
  2017{\natexlab{b}}, \aap, 598, A85

\bibitem[{{Shenar} {et~al.}(2019){Shenar}, {Sablowski}, {Hainich}, {Todt},
  {Moffat}, {Oskinova}, {Ramachandran}, {Sana}, {Sander}, {Schnurr},
  {St-Louis}, {Vanbeveren}, {G{\"o}tberg}, \& {Hamann}}]{Shenar2019}
{Shenar}, T., {Sablowski}, D.~P., {Hainich}, R., {et~al.} 2019, \aap, 627, A151

\bibitem[{{Shenar} {et~al.}(2021){Shenar}, {Sana}, {Marchant}, {Pablo},
  {Richardson}, {Moffat}, {Van Reeth}, {Barb{\'a}}, {Bowman}, {Broos},
  {Crowther}, {Clark}, {de Koter}, {de Mink}, {Dsilva}, {Gr{\"a}fener},
  {Howarth}, {Langer}, {Mahy}, {Ma{\'\i}z Apell{\'a}niz}, {Pollock},
  {Schneider}, {Townsley}, \& {Vink}}]{Shenar2021}
{Shenar}, T., {Sana}, H., {Marchant}, P., {et~al.} 2021, \aap, 650, A147

\bibitem[{{Smartt}(2009)}]{Smartt2009}
{Smartt}, S.~J. 2009, \araa, 47, 63

\bibitem[{{Strawn} {et~al.}(2023){Strawn}, {Richardson}, {Moffat}, {Ibrahim},
  {Lane}, {Pickett}, {Chen{\'e}}, {Corcoran}, {Damineli}, {Gull}, {Hillier},
  {Morris}, {Pablo}, {Thomas}, {Stevens}, {Teodoro}, \& {Weigelt}}]{Strawn2023}
{Strawn}, E., {Richardson}, N.~D., {Moffat}, A. F.~J., {et~al.} 2023, \mnras,
  519, 5882

\bibitem[{{Taylor} {et~al.}(2011){Taylor}, {Evans}, {Sana}, {Walborn}, {de
  Mink}, {Stroud}, {Alvarez-Candal}, {Barb{\'a}}, {Bestenlehner}, {Bonanos},
  {Brott}, {Crowther}, {de Koter}, {Friedrich}, {Gr{\"a}fener},
  {H{\'e}nault-Brunet}, {Herrero}, {Kaper}, {Langer}, {Lennon}, {Ma{\'\i}z
  Apell{\'a}niz}, {Markova}, {Morrell}, {Monaco}, \& {Vink}}]{Taylor2011}
{Taylor}, W.~D., {Evans}, C.~J., {Sana}, H., {et~al.} 2011, \aap, 530, L10

\bibitem[{{Tehrani} {et~al.}(2019){Tehrani}, {Crowther}, {Bestenlehner},
  {Littlefair}, {Pollock}, {Parker}, \& {Schnurr}}]{Tehrani2019}
{Tehrani}, K.~A., {Crowther}, P.~A., {Bestenlehner}, J.~M., {et~al.} 2019,
  \mnras, 484, 2692

\bibitem[{{Thomas} {et~al.}(2021){Thomas}, {Richardson}, {Eldridge},
  {Schaefer}, {Monnier}, {Sana}, {Moffat}, {Williams}, {Corcoran}, {Stevens},
  {Weigelt}, {Zainol}, {Anugu}, {Le Bouquin}, {ten Brummelaar}, {Campos},
  {Couperus}, {Davies}, {Ennis}, {Eversberg}, {Garde}, {Gardner}, {Fl{\'o}},
  {Kraus}, {Labdon}, {Lanthermann}, {Leadbeater}, {Lester}, {Maki}, {McBride},
  {Ozuyar}, {Ribeiro}, {Setterholm}, {Stober}, {Wood}, \&
  {Zurm{\"u}hl}}]{Thomas2021}
{Thomas}, J.~D., {Richardson}, N.~D., {Eldridge}, J.~J., {et~al.} 2021, \mnras,
  504, 5221

\bibitem[{{Townsley} {et~al.}(2006){Townsley}, {Broos}, {Feigelson}, {Garmire},
  \& {Getman}}]{Townsley2006}
{Townsley}, L.~K., {Broos}, P.~S., {Feigelson}, E.~D., {Garmire}, G.~P., \&
  {Getman}, K.~V. 2006, \aj, 131, 2164

\bibitem[{{Tramper} {et~al.}(2016){Tramper}, {Sana}, {Fitzsimons}, {de Koter},
  {Kaper}, {Mahy}, \& {Moffat}}]{Tramper2016}
{Tramper}, F., {Sana}, H., {Fitzsimons}, N.~E., {et~al.} 2016, \mnras, 455,
  1275

\bibitem[{{Walborn} \& {Fitzpatrick}(1990)}]{Walborn1990}
{Walborn}, N.~R. \& {Fitzpatrick}, E.~L. 1990, \pasp, 102, 379

\bibitem[{{Walborn} {et~al.}(2002){Walborn}, {Howarth}, {Lennon}, {Massey},
  {Oey}, {Moffat}, {Skalkowski}, {Morrell}, {Drissen}, \&
  {Parker}}]{Walborn2002}
{Walborn}, N.~R., {Howarth}, I.~D., {Lennon}, D.~J., {et~al.} 2002, \aj, 123,
  2754

\bibitem[{{Weigelt} \& {Baier}(1985)}]{Weigelt1985}
{Weigelt}, G. \& {Baier}, G. 1985, \aap, 150, L18

\bibitem[{{Woosley} {et~al.}(2007){Woosley}, {Blinnikov}, \&
  {Heger}}]{Woosley2007}
{Woosley}, S.~E., {Blinnikov}, S., \& {Heger}, A. 2007, \nat, 450, 390

\bibitem[{{Zucker} \& {Mazeh}(1994)}]{Zucker1994}
{Zucker}, S. \& {Mazeh}, T. 1994, \apj, 420, 806

\end{thebibliography}

\begin{appendix}

\section{Observation log and RV measurements}
\label{appendix:log}

Tables\,\ref{tab:RVs} and \ref{tab:RVsC} compile the RV measurements for a1, a2, a3, and c using the STIS/HST data and the ARGUS/FLAMES data, respectively. Table\,\ref{tab:EWs} compiles the EWs of several lines in the STIS/HST data, while Table\,\ref{tab:FWHM} provides the FWHM of the \NIV~and \HeII~lines for the STIS/HST dataset.

\begin{table*}
\centering
\caption{RVs (in ${\rm km}\,{\rm s}^{-1}$) for a1, a2, a3, and c, as derived from the STIS/HST data}
\begin{tabular}{lccccccc}
\hline
Object & MJD & S/N & RV (N\,{\sc iv}\,$4058$) & RV (N\,{\sc v}\,$\lambda \lambda 4603, 4621$) & RV (He\,{\sc ii}\,$\lambda 4686$) & RV(N\,{\sc v}\,$\lambda 4945$) \\ 
\hline
a1 & 56023.44 & 36 & 354.4 $\pm$ 6.2 & 327.1 $\pm$ 11.2 & 307.6 $\pm$ 6.2 &  -  \\ 
 & 58936.20 & 36 & 335.0 $\pm$ 7.3 & 310.0 $\pm$ 9.1 & 285.4 $\pm$ 6.0 & 346.7 $\pm$ 10.3 \\ 
 & 59120.06 & 47 & 336.1 $\pm$ 9.2 & 325.9 $\pm$ 10.0 & 319.6 $\pm$ 5.5 & 348.3 $\pm$ 16.4 \\ 
 & 59471.74 & 34 & 340.6 $\pm$ 8.7 & 334.5 $\pm$ 9.6 & 296.0 $\pm$ 5.3 & 352.1 $\pm$ 13.9 \\ 
\rule{0pt}{15pt}
a2 & 56023.84 & 134 & 350.9 $\pm$ 6.5 & 335.2 $\pm$ 8.4 & 220.3 $\pm$ 4.0 &  -  \\ 
 & 58936.20 & 29 & 341.4 $\pm$ 10.3 & 307.2 $\pm$ 9.5 & 239.0 $\pm$ 5.0 & 356.8 $\pm$ 14.7 \\ 
 & 59120.07 & 64 & 337.7 $\pm$ 8.5 & 320.1 $\pm$ 9.9 & 215.5 $\pm$ 3.7 & 352.5 $\pm$ 7.8 \\ 
 & 59471.73 & 38 & 335.5 $\pm$ 7.1 & 328.1 $\pm$ 10.0 & 240.2 $\pm$ 4.0 & 352.3 $\pm$ 8.6 \\ 
\rule{0pt}{15pt}
a3 & 56023.42 & 33 & 341.3 $\pm$ 11.3 & 301.9 $\pm$ 11.4 & 231.1 $\pm$ 4.3 &  -  \\ 
 & 58994.06 & 21 & 347.3 $\pm$ 9.4 & 276.2 $\pm$ 9.4 & 234.5 $\pm$ 5.2 & 334.3 $\pm$ 16.5 \\ 
 & 59179.31 & 19 & 346.9 $\pm$ 9.9 & 270.6 $\pm$ 10.4 & 237.4 $\pm$ 5.8 & 359.1 $\pm$ 15.0 \\ 
 & 59348.22 & 42 & 341.0 $\pm$ 9.0 & 261.5 $\pm$ 9.2 & 212.8 $\pm$ 4.7 & 315.7 $\pm$ 23.7 \\ 
\rule{0pt}{15pt}
c & 58994.06 & 14 & 393.1 $\pm$ 19.1 & 290.0 $\pm$ 20.0 & 333.7 $\pm$ 8.5 & 393.4 $\pm$ 19.4 \\ 
 & 59179.31 & 19 & 347.4 $\pm$ 14.3 & 343.0 $\pm$ 15.0 & 275.5 $\pm$ 7.4 & 373.4 $\pm$ 19.3 \\ 
 & 59348.22 & 14 & 326.4 $\pm$ 25.5 & 260.4 $\pm$ 16.5 & 244.1 $\pm$ 7.9 & 303.8 $\pm$ 43.1 \\ 
\hline
\end{tabular}
\label{tab:RVs}
\end{table*}

\begin{table*}
\centering
\caption{EWs for diagnostic lines of a1, a2, a3, and c, as derived from the STIS/HST data (in units of $\AA$) $^{\rm (a)}$ }
\begin{tabular}{lcccccc}
\hline
Object & MJD &  EW (N\,{\sc iv}\,$4058$) & EW (N\,{\sc v}\,$\lambda \lambda 4603, 4621$) & EW (He\,{\sc ii}\,$\lambda 4686$) & EW(N\,{\sc v}\,$\lambda 4945$) \\ 
\hline
a1 & 56023.44 & -2.1$\pm 0.4$ & 0.5$\pm 0.5$ & -30.8$\pm 0.9$ &  -  \\ 
 & 58936.20 & -2.2$\pm 0.5$ & 0.1$\pm 0.6$ & -38.5$\pm 2.4$ & -0.22$\pm 0.08$ \\ 
 & 59120.06 & -2.4$\pm 0.7$ & -0.1$\pm 0.7$ & -35.0$\pm 1.5$ & -0.15$\pm 0.05$ \\ 
 & 59471.74 & -2.8$\pm 1.2$ & 0.8$\pm 0.9$ & -36.6$\pm 2.5$ & -0.07$\pm 0.08$ \\ 
\rule{0pt}{15pt}
a2 & 56023.84 & -1.7$\pm 0.4$ & 0.5$\pm 0.4$ & -33.8$\pm 1.8$ &  -  \\ 
 & 58936.20 & -2.4$\pm 0.7$ & 0.4$\pm 0.7$ & -40.8$\pm 1.4$ & -0.33$\pm 0.09$ \\ 
 & 59120.07 & -2.2$\pm 0.6$ & 0.1$\pm 0.6$ & -37.5$\pm 1.3$ & -0.42$\pm 0.10$ \\ 
 & 59471.73 & -2.1$\pm 0.7$ & 0.8$\pm 0.6$ & -37.9$\pm 2.4$ & -0.22$\pm 0.08$ \\ 
\rule{0pt}{15pt}
a3 & 56023.42 & -3.0$\pm 1.0$ & 0.7$\pm 0.5$ & -52.8$\pm 1.6$ &  -  \\ 
 & 58994.06 & -3.7$\pm 1.6$ & 0.5$\pm 1.5$ & -58.3$\pm 4.5$ & -0.41$\pm 0.09$ \\ 
 & 59179.31 & -3.0$\pm 1.3$ & 1.1$\pm 2.1$ & -55.1$\pm 6.1$ & -0.40$\pm 0.12$ \\ 
 & 59348.22 & -3.0$\pm 1.1$ & -1.0$\pm 1.5$ & -58.0$\pm 5.4$ & -0.68$\pm 0.13$ \\ 
\rule{0pt}{15pt}
c & 58994.06 & -2.4$\pm 1.8$ & 1.3$\pm 2.5$ & -48.0$\pm 6.3$ & -0.12$\pm 0.08$ \\ 
 & 59179.31 & -2.5$\pm 1.0$ & -2.3$\pm 3.0$ & -55.9$\pm 4.7$ & -0.43$\pm 0.08$ \\ 
 & 59348.22 & -2.9$\pm 1.3$ & -0.5$\pm 2.7$ & -53.2$\pm 5.1$ & -0.24$\pm 0.11$ \\ 
\hline
\end{tabular}
\label{tab:EWs}
\tablefoot{$^{\rm (a)}$ Errors are computed via equation (A9) in \citet{Chalabaev1983} and represent upper limits on the statistical errors.}
\end{table*}

\begin{table}
\centering
\caption{FWHMs (in \AA)  of the \NIV~and \HeII~lines in the STIS/HST datasets}
\begin{tabular}{lccc}
\hline
Object & MJD & FWHM (\NIV) & FWHM (\HeII) \\ 
\hline
a1 & 58936.20 & 6.33$\pm 0.51$ & 27.26$\pm 0.37$ \\ 
 & 59471.74 & 7.18$\pm 0.83$ & 24.77$\pm 0.46$ \\ 
 & 59120.06 & 6.80$\pm 0.57$ & 25.69$\pm 0.40$ \\ 
 & 56023.44 & 5.81$\pm 0.37$ & 27.60$\pm 0.37$ \\ 
\rule{0pt}{15pt}
a2 & 58936.20 & 7.60$\pm 0.94$ & 25.83$\pm 0.42$ \\ 
 & 59120.07 & 6.42$\pm 0.51$ & 27.13$\pm 0.52$ \\ 
 & 56023.84 & 5.65$\pm 0.39$ & 24.90$\pm 0.37$ \\ 
 & 59471.73 & 6.37$\pm 0.59$ & 27.56$\pm 0.39$ \\ 
\rule{0pt}{15pt}
a3 & 58994.06 & 6.74$\pm 0.84$ & 26.55$\pm 0.52$ \\ 
 & 59348.22 & 6.67$\pm 0.75$ & 26.59$\pm 0.56$ \\ 
 & 56023.42 & 7.58$\pm 0.41$ & 27.79$\pm 1.28$ \\ 
 & 59179.31 & 6.43$\pm 0.67$ & 26.77$\pm 0.52$ \\ 
\rule{0pt}{15pt}
c & 58994.06 & 7.66$\pm 1.92$ & 24.90$\pm 0.67$ \\ 
 & 59179.31 & 6.50$\pm 1.26$ & 26.39$\pm 0.58$ \\ 
 & 59348.22 & 8.24$\pm 1.41$ & 26.05$\pm 0.62$ \\ 
\hline
\end{tabular}
\label{tab:FWHM}
\end{table}

\begin{table}
\centering
\caption{RVs (in ${\rm km}\,{\rm s}^{-1}$) for R\,136~c, as derived from the UVES and FLAMES data}
\begin{tabular}{lccc}
\hline
Instrument & MJD & S/N & RV (N\,{\sc iv}\,$4058$) \\ 
\hline
UVES & 52176.296 & 40 & 298.5 $\pm$ 2.2 \\ 
FLAMES & 54761.217 & 86 & 327.7 $\pm$ 5.9 \\ 
FLAMES & 54761.224 & 46 & 339.5 $\pm$ 5.9 \\ 
FLAMES & 54761.230 & 57 & 339.6 $\pm$ 6.6 \\ 
FLAMES & 54761.237 & 43 & 332.1 $\pm$ 5.6 \\ 
FLAMES & 54761.244 & 97 & 343.2 $\pm$ 6.0 \\ 
FLAMES & 54761.251 & 76 & 335.1 $\pm$ 5.7 \\ 
FLAMES & 54761.267 & 35 & 340.4 $\pm$ 6.4 \\ 
FLAMES & 54761.273 & 117 & 339.7 $\pm$ 6.2 \\ 
FLAMES & 54761.280 & 51 & 340.6 $\pm$ 5.5 \\ 
FLAMES & 54761.287 & 39 & 342.5 $\pm$ 7.8 \\ 
FLAMES & 54761.293 & 59 & 343.6 $\pm$ 7.2 \\ 
FLAMES & 54761.300 & 75 & 334.1 $\pm$ 7.0 \\ 
FLAMES & 54767.263 & 93 & 321.5 $\pm$ 8.1 \\ 
FLAMES & 54767.270 & 76 & 320.2 $\pm$ 8.6 \\ 
FLAMES & 54767.277 & 79 & 324.1 $\pm$ 9.4 \\ 
FLAMES & 54767.283 & 110 & 316.8 $\pm$ 8.0 \\ 
FLAMES & 54767.290 & 67 & 328.4 $\pm$ 7.9 \\ 
FLAMES & 54767.297 & 23 & 306.2 $\pm$ 10.8 \\ 
FLAMES & 54845.157 & 59 & 331.4 $\pm$ 10.1 \\ 
FLAMES & 54845.163 & 109 & 319.4 $\pm$ 10.2 \\ 
FLAMES & 54845.170 & 69 & 333.6 $\pm$ 10.9 \\ 
FLAMES & 54845.177 & 70 & 323.8 $\pm$ 9.7 \\ 
FLAMES & 54845.183 & 57 & 321.9 $\pm$ 9.8 \\ 
FLAMES & 54845.190 & 36 & 324.4 $\pm$ 11.8 \\ 
FLAMES & 54876.111 & 75 & 272.5 $\pm$ 11.9 \\ 
FLAMES & 54876.118 & 65 & 276.9 $\pm$ 11.3 \\ 
FLAMES & 54876.124 & 134 & 269.6 $\pm$ 10.2 \\ 
FLAMES & 54876.131 & 60 & 267.8 $\pm$ 9.8 \\ 
FLAMES & 54876.138 & 38 & 272.2 $\pm$ 10.8 \\ 
FLAMES & 54876.144 & 122 & 280.8 $\pm$ 10.2 \\ 
FLAMES & 55173.288 & 210 & 328.7 $\pm$ 10.2 \\ 
FLAMES & 55173.310 & 144 & 326.0 $\pm$ 11.1 \\ 
FLAMES & 55178.145 & 183 & 335.9 $\pm$ 7.4 \\ 
FLAMES & 55178.167 & 95 & 338.3 $\pm$ 6.8 \\ 
\hline
\end{tabular}
\label{tab:RVsC}
\end{table}

\section{Detection probabilities for highly eccentric binaries}
\label{appendix:ecc}

\begin{figure}
\centering
\includegraphics[width=.5\textwidth]{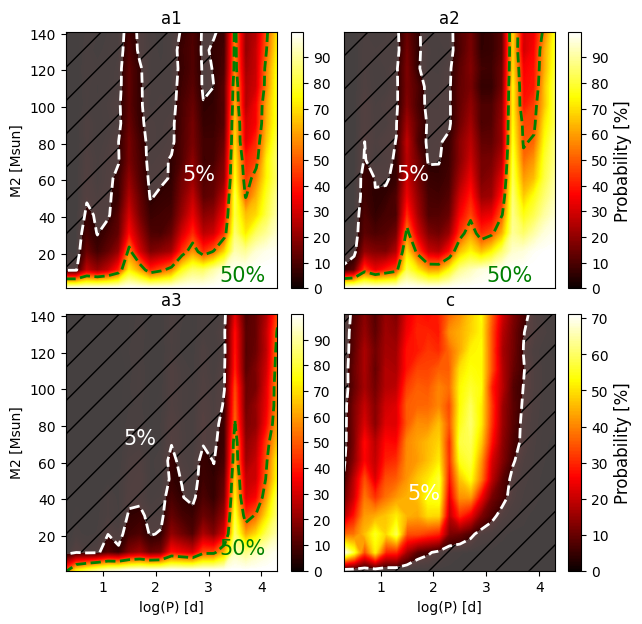}
\caption{As Fig.\,\ref{fig:prob} but for an underlying Gaussian eccentricity distribution with a mean of $<e> = 0.8$ and standard deviation of 0.1  } 
\label{fig:prob_ecc}
\end{figure}

Some known massive binaries in the LMC exhibit high eccentricities (e.g., \object{R\,145}, $e=0.79$, \citealt{Shenar2017}; \object{R\,144}, $e=0.56$, \citealt{Shenar2021}; \object{Mk\,34}, $e=0.76$, \citealt{Tehrani2019}), while others exhibit more moderate eccentricities (e.g., \object{Mk\,33Na}, $e = 0.33$, \citealt{Bestenlehner2022};  \object{R~139}, $e = 0.38$, \citealt{Taylor2011, Mahy2020a}). To explore the impact of potential high eccentricity in our targets,  we repeat the exercise performed in Sect.\,\ref{sec:discussion} for a Gaussian eccentricity distribution with a mean of $\langle e \rangle = 0.8$ and a standard deviation of 0.1 (Fig.\,\ref{fig:prob_ecc}). As could be anticipated, the detection probability drops, though the exclusion domains are still comparable. Only highly eccentric binaries ($e > 0.9$) would have an appreciable likelihood to evade detection even at shorter ($\lesssim 100\,$d) orbital periods.  More epochs would certainly improve the detection probability of high eccentricity binaries.

\end{appendix}

\end{document}